\documentstyle[12pt]{article}

\newtheorem{defi}{Definition}
\newcommand{\bdf}{\begin{defi}}
\newcommand{\edf}{\end{defi}}
\newtheorem{prop}{Proposition}[section]
\newcommand{\bpr}{\begin{prop}}
\newcommand{\epr}{\end{prop}}
\newtheorem{theo}{Theorem}
\newcommand{\bth}{\begin{theo}}
\newcommand{\eth}{\end{theo}}
\newtheorem{coro}{Corollary}
\newcommand{\bcr}{\begin{coro}}
\newcommand{\ecr}{\end{coro}}
\newtheorem{lemma}{Lemma}[section]
\newcommand{\blm}{\begin{lemma}}
\newcommand{\elm}{\end{lemma}}
\newtheorem{example}{Example}
\newcommand{\bex}{\begin{example}}
\newcommand{\eex}{\end{example}}
\newcommand{\rf}[1]{{\rm (\ref{#1})}}

\def\bce{\begin{center}}
\def\ece{\end{center}}
\def\beq{\begin{eqnarray}}
\def\eeq{\end{eqnarray}}
\def\ben{\begin{enumerate}}
\def\een{\end{enumerate}}

\def\ni{\noindent}
\def\nnb{\nonumber}

\newcommand{\pr}{\noindent {\bf Proof:} \hspace{3mm}}

\def\tg{{\tilde g}}
\def\btg{\tg}
\def\ab{\alpha \beta}   
\def\lm{\lambda \mu}    
\def\ol{\Omega \Lambda}   

\def\bsigma{{\bar \sigma}}
\def\brho{{\bar \rho}}

\def\re{\varphi^{*} }
\def\rea{\re_t }

\def\xiv{\vec \xi}          
\def\xiva{\xiv_{\!_{A} }}  
\def\xivb{\xiv_{\!_{B} }}  

\def\lie{{\cal L}({\scriptstyle \xiv}) }  

\def\lh{\hbox{{\boldmath \hbox{$ \ell $}}}}
\def\lv{\vec \ell }    
\def\mh{\hbox{{\boldmath \hbox{$ m $}}}}
\def\kh{\hbox{{\boldmath \hbox{$ k $}}}}
\def\bkh{\hbox{{\boldmath \hbox{$ \bar k $}}}}
\def\ph{\hbox{{\boldmath \hbox{$ p $}}}}  
\def\qh{\hbox{{\boldmath \hbox{$ q $}}}}  
\def\rh{\hbox{{\boldmath \hbox{$ r $}}}}   
\def\uh{\hbox{{\boldmath \hbox{$ u $}}}}   
\def\nh{\hbox{{\boldmath \hbox{$ n $}}}}  
\def\mh{\hbox{{\boldmath \hbox{$ m $}}}} 
\def\ll{\lh \otimes \lh }        
\def\mm{\mh \otimes \mh }    
\def\pp{\ph \otimes \ph }         
\def\qq{\qh \otimes \qh }         
\def\uu{\uh \otimes \uh }         
\def\nn{\nh \otimes \nh }         
\def\slm{\lh \otimes \mh + \mh \otimes \lh}
\def\spq{\ph \otimes \qh + \qh \otimes \ph}
\def\sun{\uh \otimes \nh + \nh \otimes \uh}

\def\of{\hbox{{\boldmath \hbox{$ \Theta $}}} }
\def\th{\hbox{{\boldmath \hbox{$ \Theta $}}} }
\def\ofo{ {\of}^{\Omega}} 
\def\ofl{ {\of}^{\Lambda}} 
\def\tol{\of\sp{\Omega} \otimes \of\sp{\Lambda} }

\def\dx{\hbox{{\boldmath \hbox{$ d $}}} }
\def\cb{ \{ {\of}^{\Omega} \} } 
\def\fa{(Q_{t})_{\Omega\, \Lambda}} 
\def\qx{q_{\xi}}    
\def\qxb{(q_{\xi})_{\ol}} 
\def\bg{{g}}                      
\def\sign{{\rm sign}}  

\def\ks{Kerr-Schild }
\def\gks{GKS }
\def\st{spacetime }

\def\N{\hfill{}\rule{2mm}{2mm}}

\begin{document}
\title{\ks and generalized metric motions}
\author{Sergi R. Hildebrandt \\
{\small Instituto de Ciencias del Espacio (CSIC) \&}\cr
{\small Institut d'Estudis Espacials de Catalunya (IEEC/CSIC)}\cr
{\small Edifici Nexus, Gran Capit\`a 2-4, 08034 Barcelona, Spain}\cr
{\small hildebrandt@ieec.fcr.es, http://www.ieec.fcr.es/cosmo-www/}}
\maketitle
\date{}
\begin{abstract}
In this work we study in detail new kinds of motions of the metric tensor. The work 
is divided into two main parts.
In the first part we study the general existence of \ks
motions ---a recently introduced metric motion. We show that generically,
\ks motions give rise to finite dimensional Lie algebras and are {\em
isometrizable}, i.e., they are in a one-to-one correspondence with a subset of
isometries of a (usually different) spacetime. This is similar to conformal
motions. There are however some exceptions that yield infinite dimensional
algebras in any dimension of the manifold. We also show that \ks motions may
be interpreted as some kind of metric symmetries in the sense of having
associated some geometrical invariants. In the second part, we suggest a
scheme able to cope with other new candidates of metric motions from a
geometrical viewpoint. We solve a set of new candidates which may be
interpreted as the seeds of further developments and relate them with known
methods of finding new solutions to Einstein's field equations. The
results are similar to those of \ks motions, yet a richer algebraical
structure appears. In conclusion, even though several points still remain
open, the wealth of results shows that the proposed concept of {\em
generalized metric motions} is meaningful and likely to have a spin-off in
gravitational physics.We end by listing and analyzing some of those open
points.
\end{abstract}

\section{Introduction}
\label{intro}
This work deals with motions
of the metric tensor of a Riemannian manifold ---see e.g.,~\cite{hall962} for
a review--- that,
in physical terms, is to be assimilated with
spacetime.  In particular, we will give some examples of how one can
extend the number of symmetries of the metric tensor
beyond the two cases that have been mostly considered
in the literature, namely isometries and conformal sym\-metries.

We will work along two lines. Continuous groups of Generalized \ks (GKS)
transformations, or simply \ks motions, have been presented in \cite{ksvf}
(hereafter referred to as I). In the first part of this work we will carry
out a detailed study of this particular new metric motion (see also below).
This part aims at showing how the usual study of conformal or isometric
symmetries can be extended to other cases having a role in gravitation. On
the other side, the aim of the second part is  to pose and develop
a framework to deal with some general situations, since
one expects that many achievements of differential geometry on
Riemannian spaces have a physical spin-off. Let us now give a more detailed
account of each part.

In I some general  properties and explicit examples of \ks motions were
presented. In this work we shall address the question of their general
existence in any \st of arbitrary dimension, i.e., a detailed study of the
solutions of their associated differential equations, the so-called \ks
equations (see following section). The main results are that \ks motions are
mainly divided into three families depending on the kinematical properties
of the deformation direction
$ \lv $. Moreover,
we will show that generically they give rise to Lie algebras of finite
dimensional character as happens with isometries or conformal motions,
although the existence of infinite dimensional Lie algebras is possible. We
also show that generically \ks motions are ``{\em isometrizable}'', i.e., they are equivalent to a subset of the
isometries of a ---usually different--- spacetime. This solves a question
raised in I.

In fact this scheme could in principle be extended to other new
cases of metric motions, see below. However,
besides comparing the contents of
this chapter with other new proposals, it is worth
relating it with current studies in other types of motions, namely
collineations, see e.g.,
\cite{archel}--\cite{hs} for curvature collineations,
\cite{tsm}--\cite{qz} (and also~\cite{hdc}) for Ricci collineations,
\cite{marteens}--\cite{hl} for affine
collineations and holonomy theory, \cite{weylp}--\cite{hk} for projective
collineations ---see also \cite{hall002}---, and
\cite{schouten}--\cite{kld} for a general  presentation and some
fundamental results.
We do not dare to follow
this way here for the sake of brevity. With this in hand, it should
be possible to place the object under study within the
family of metric motions, or even
within general motions, see also \cite{hall962,hall96,hall98}.

Afterwards, in section~\ref{s-gmm}, we introduce a
framework for studying ``the family'' of metric symmetries in general. This
framework is
then applied to some specific cases of physical interest.
The results show that the mathematical structure of its
solutions utterly surpasses the well known results for isometric and
conformal motions, or those given in Sect.~\ref{s-ksm} for the \ks case, yet
a similar status as that of these two fundamental
metric symmetries is still to be worked out.

Moreover, one recovers, as particular cases, isometric and conformal
symmetries, as well as the \ks case, and this consistency gives us a good
control on the computations carried out, which are sometimes involved.
One may ask which is the interest of this all. This question has not a
simple answer. From our viewpoint the interest is twofold.
On the one hand, there is the purely mathematical interest,
of finding a useful way to characterize
the most fundamental symmetries of a Riemannian manifold
---the symmetries of the metric tensor.
The scheme here developed proves to be worth
towards such goal. However, for a physicist, it is impossible
to detach this question from another one on the practical
applications that
these ``generalized metric symmetries'' can contribute in this domain.
Thus, right from the first step, in which the example
of a new metric symmetry (the ``Kerr-Schild case'') is given,
to the final implementation of the very general mathematical framework, it
is physics which sets the pace.
It is our opinion that this feedback between
their mathematical and their physical features will govern the
future study of new possible metric motions.
Some other physical applications will be given elsewhere~\cite{
pakss}.
Finally, let us add that we have provided some
examples, also with the intention of giving practical tools for
solving other situations.

Section~\ref{s-ksm} deals with \ks motions. We begin recalling
a definition and two results of I which are necessary for the rest of the
work. In Sect.~\ref{ss-ngk} the case of non-geodesic \ks motions is solved
and their isometrization is proven. In Sect.~\ref{ss-gldn0} we solve the
case of geodesic \ks motions with a deformation direction that satisfies
some particular conditions on its kinematical properties. This, we will
show, covers almost any case of geodesic \ks motions. We also prove a
similar isotropization as in previous case. In Sect.~\ref{sss-kskn} we give
the general solution of \ks motions for Kerr-Newman spacetimes and their
principal null directions. In Sect.~\ref{s-gld0} we address the remaining
cases. We show that there is {\em no general solution} analogue to previous
cases and in this case the associated Lie algebras may become infinite
dimensional in any dimension of the manifold (in oposition to isometric or
conformal motions, where only the latter has Lie algebras of infinite
dimensional character and only in the case of a two-dimensional manifold).
Moreover, we show that their existence is indeed strongly restricted and we
give the main features of these cases. In Sect.~\ref{sss-sld0} we study in
detail the existence of this possibility under some physically interesting
conditions. In Sect.~\ref{ss-sksm} we give a brief summary of the \ks part.
In Sect.~\ref{s-gmm} we deal with the definition of (continuous groups of)
{\em generalized metric motions}. We begin with some motivations in
Sects.~\ref{ss-rcpa} and~\ref{ss-pfgm}. Our choice is presented in
Sect.~\ref{ss-dgmg} in Def.~\ref{def-ms}. In Sect.~\ref{ss-ivmg} we write
down its differential version. In Sect.~\ref{s-mmgd} the framework is
implemented to the study of most relevant cases, i.e., metric motions
generated by 1-dimensional subspaces. In particular, we study in detail the
existence of such motions, their interrelation and, particularly, in
Ex.~\ref{exlm} we present the general solution for the case of two
covariantly constant 1-form fields. In Sect.~\ref{ss-acg} we address some
questions regarding the addition of a conformal motion to a given metric
motion. In Sect.~\ref{ss-rpa} we give a list of further points that in our
opinion seem worth for a deeper study in the future. We finish in
Sect.~\ref{s-con} giving the general conclusions. In App.~\ref{app-a} a
survey of useful formulae when dealing with \ks motions is given. In
App.~\ref{app-b} first steps towards the the integrability equations for a {\em generalized metric
motion} under our scheme, including the \ks case, are written down. In
App.~\ref{app-c} we summarize some features of  metric motions generated by
a spacelike and a timelike 1-form fields. Finally,
in App.~\ref{app-d} some hints towards a {\em complete}
(explicit) resolution of \ks motions in flat \st is given.

The conventions and notation used throughout this work are the following. $ (V_n, g) $ denotes a smooth connected Hausdorff $ n $-dimensional manifold admitting a smooth (Lorentz) metric $ g $. We will use the signature
(--1,1,\dots,1), although Euclidean or other types of signatures can be
considered. Greek  indices run from 0 to
$ n-1 $, whereas Latin indices run from 1 to $ n-1$. The
tensor product is denoted by $ \otimes $.
1-form fields, that is 1-covariant
tensor fields, are denoted in boldface, e.g., $ \lh $, $ \uh $, $ \ph $,
\ldots $ \, $Symmetrization and antisymmetrization on a pair of indices is
defined by $ A_{(\ab)} \equiv (1/2) (A_{\ab} + A_{\beta \alpha}) $, and $
A_{[\ab]} \equiv (1/2) (A_{\ab} - A_{\beta \alpha}) $. The Riemann, or
curvature, tensor is defined as $ R^{\alpha}_{\, \beta \gamma \delta} \equiv
\partial_{\gamma} \Gamma^{\alpha}_{\beta \delta} -
  \partial_{\delta} \Gamma^{\alpha}_{\beta \gamma} +
\Gamma_{\gamma \lambda}^{\alpha} \Gamma^{\lambda}_{\beta \delta} -
\Gamma_{\delta \lambda}^{\alpha} \Gamma_{\beta \gamma}^{\lambda} $. The
Ricci tensor is defined by
$ R_{\ab} \equiv R^{\lambda}_{\; \alpha \lambda \beta} $.
Other symbols used throughout this work are:
$ D \equiv \ell^{\lambda}\nabla_{\lambda} $, $ \theta \equiv (1/2)
\nabla_{\lambda} \ell^{\lambda} $, $ \bar R \equiv 2 \Phi_{00}
\equiv R_{\sigma \lambda} \ell^{\sigma} \ell^{\lambda} $,
$ \varsigma^2 \equiv (1/2) \nabla_{(\sigma}\ell_{\lambda)}
\nabla^{\sigma} \ell^{\lambda} - \theta^2 $, $ \varpi^2
\equiv (1/2) \nabla_{[\sigma}\ell_{\lambda]}\nabla^{\sigma}
\ell^{\lambda}$. $ \Psi_{0} \equiv C_{\lambda \sigma \mu \rho}
\ell^{\lambda}
k^{\sigma} \ell^{\mu} k^{\rho} $, with $ \kh $ a null
complex-valued 1-form orthogonal to $ \lh $. If~$ \lh $ is geodesic, $ D \lh
\equiv M \lh $,
where $ M $ is a function.
Throughout this work we will edal with {\em local} 1-parameter groups of {\em local} motions acting upon the metric tensor and we shall simplt call them ``metric motions''. A ``motion'' being itself a (smooth) diffeomorphism. Finally, the end of a proof or its absence is marked by
$ \; \rule{2.5mm}{2.5mm}$.
\section{ \ks motions}
\label{s-ksm}
We begin by giving some basic concepts of \ks motions (proofs and details 
are
available in I)
\bdf[\ks vector fields]
\label{defksv}
Any solution $\xiv$ of the equations (hereafter called \ks equations)
\beq
\lie g = 2 h \, \lh \otimes \lh,
\label{eq-h}\\
\lie \lh = m \, \lh \ \label{eq-sta}
\eeq
where $h$ and $m$ are two functions over $ V_n $ and $ \lh $ is a null
1-form field will be called a
{\em \ks vector field (KSVF)} with respect to $\lh$. The functions $h$ and
$m$ are the {\em gauges} of the metric $g$ and of $\lh$, respectively, and $
\lh $ is called the {\em deformation direction}.
\edf
It is easy to show that {\em only} the direction of $ \lh $ is of relevance
since $ \lh $ is null, i.e., for any oher $ \lh^{\prime} = A \lh $, with $ A
\neq 0 $ an arbitrary function, the KSVFs with respect to $ \lh^{\prime} $
are the same as with respect to $ \lh $. Therefore, any general result
regarding \ks equations has to
include this basic property.

Other results are
\bpr
\label{prop-two}
Two metrics linked by a GKS relation,
$ {\bar g} = g + 2 H \lh \otimes \lh$,
admit the same KSVFs with respect to $\lh$.
\label{same}
\N
\epr
The possibility of the infinite dimensional character of some Lie algebras
is easily seen from (a slighty variant of theorem 2 in I)
\bth
\label{open}
For any $ \lh $ such that $ \nabla \lh = b \lh \otimes
\lh $ where $ {\bf b} $ is some function, $ \xiv = \rho \, \lv $ with $
d\rho \ne 0 $
is a KSVF with respect to the same direction $ \lh $ if
and only if the functions $\rho$ are those of
the ring generating $\lh$, that is to say, such that
$ \lh \wedge d\rho = 0$. \N
\eth
In the sequel, the metric
tensor $ g $ and the direction of the deformation $ \lh $
are to be considered as data in \ks equations.
On the other hand, $ h $ and $ m $ are unknown $ C^{\infty} $
functions and $ \xiv $, a KSVF, is also an unknown of the problem.

Let us start with the first case.

\subsection{Non-geodesic \ks motions}
\label{ss-ngk}
We will call {\em non-geodesic \ks motions} whenever the vector field $ \lv
$ is not geodesic. We recall that a null vector field on $ V_n $ is
non-geodesic if and only if $ a^2 (\equiv g_{\lm} a^{\lambda} a^{\mu})  \ne
0 $, where $ \vec a $ is the four-acceleration vector associated with $ \lv
$, defined by $ {\vec a} \equiv D \lv $. In this case we have,
\bth
\label{ksng}
The system of non-geodesic \ks motions is closed.
\eth
\pr From the expression of $ \lie R_{\ab} $ given in App.~\ref{app-a} and
the null character of $ \lh $ one gets
$ \ell^{\lambda} \ell^{\mu} \lie R_{\lambda \mu} = - 2 h a^2 $.
Whence $ h =- (\ell^{\lambda} \ell^{\mu}/2a^2)
\lie R_{\lambda \mu} $. Thus $ h $ is isolated
in terms of data, and the unknowns $ \xi_{\alpha} $, so that
the system is closed.\footnote{That is, one could now rewrite
Eqs.~\rf{eq-h} in an
explicit normal form for $ h $, $ \xi_{\alpha} $ and
$ \xi_{\ab} $,
as it is done in the conformal case ---see e.g., \cite{yano}.
However, due to following results, this is actually
secondary, because the \ks equations will be linked
with Killing equations, which have a simple and well known
normal system. This result will also be applied in the following section.} \N

Substituting the expression for $ h $ into \ks equations one gets
\bth[Isometrization$\!$ of non-geodesic$\!$ \ks motions]
\label{isomksng}
The KSVFs \allowbreak of a non-geodesic \ks motion are
the Killing vector fields of $ (V_n,\gamma) $, where
\beq
\gamma \equiv g + ({\bar R}/a^2)\ll , \nnb
\eeq
{\em restricted} by Eqs.~\rf{eq-sta}.
\eth
\pr First we compute $ \lie a^2 $. To do this, we first
use the formula~\rf{app-1} of App.~\ref{app-a}
and  that $ a_{\mu} \ell^{\mu} $ vanishes 
for any null vector field
 to obtain $ \lie a^{\rho} =
\lie (\ell^{\sigma} \nabla_{\sigma} \ell^{\rho} )=
(\lie \ell^{\sigma})\nabla_{\sigma}\ell^{\rho} +
D(\lie \ell^{\rho}) $. Recalling Eq.~\rf{app-2}
we readily get $ \lie {\vec a} = 2 m {\vec a} + (Dm) \lv $.
Therefore, $ \lie a^2 = \lie (g_{\lm} a^{\lambda} a^{\mu}) = 2 h
\ell_{\lambda} \ell_{\mu} a^{\lambda} a^{\mu} + 2 a_{\rho} \lie a^{\rho} =
4m a^2 $.

Now, the expression of $ h $ found in the proof of
theorem~\ref{ksng} can be rewritten as
$ h = \lie (-{\bar R}/2a^2)  - m{\bar R}/a^2 $,
where we have put $ {\bar R} \equiv R_{\mu \nu} \ell^{\mu} \ell^{\nu} $.
Hence, we can write $ 2h\ll $ as
$ \lie (-{\bar R}\ll/a^2) $, which is the key result.
On the other hand, the initial problem is completely
characterized by $ \lie g = 2 h \ll $ and $ \lie \lh = m \lh $.
Taking into account the expression obtained for $ h $, for
a non-geodesic $ \lh $, the system is totally equivalent to
$ \lie \gamma = 0 $ and $ \lie \lh = m \lh $, with
$ \gamma \equiv g + ({\bar R}/a^2) \ll $. Notice
that the expression of $ \gamma $ is independent
of the parametrization of $ \lv $:
if $ \lv \to A \lv $, $ \vec a \to A^2 {\vec a} + A(DA)\, \lv $ and
$ a^2 \to A^4 a^2 $. Therefore $ \gamma $ remains invariant and the set of
KSVFs is the same, as mentioned before.

Furthermore, since $ \gamma $ is linked by a
\gks relation with $ g $, usual results on \gks
relations ---see e.g., \cite{kramer}--- assure that $ \gamma $
is non-degenerated ($ \det \gamma = \det g \neq 0 $),
and may be reinterpreted as another metric tensor.
Besides this, we also have that $ \lh $ is a
non-geodesic null 1-form in $ (V_n, \gamma) $
(we use that $ a^2 $ is invariant if
two spacetimes are linked by a \gks relation).
In conclusion, the \ks motions for a non-geodesic
$ \lh $ are equivalent to the set of isometries of
$ (V_n,\gamma) $ that satisfy Eqs.~\rf{eq-sta} or equivalently that commute
with the direction
$ \lv $ according to Eq.~\rf{app-1}, i.e.,
$ [\xiv, \lv] = m \lv $. \N

Moreover, the theorem above asserts, from a structural point
of view, that {\em all} non-geodesic \ks motions
are ``isometrizable'' under an appropriate GKS
relation with the initial spacetime. We thus have that non-geodesic \ks
motions are affine motions with respect the Levi-Civitta connection of $
\gamma $ and are hence {\em linearizable}  \cite{hall962}.
This by itself constitutes an extension of the works~\cite{yano,bilyalov}--\cite{hst} to the case of \ks motions. These
authors had considered the analogous problem in the case of
conformal symmetries, i.e., when a group of conformal
symmetries may become a group of isometries of a
certain conformally related spacetime. Our result has
an extra bonus, namely, it turns out to be independent
of the particular properties of the Lie algebra.
Besides that, it is sometimes a useful tool for finding the
set of KSVFs of a given problem.
For instance, we have
\bpr
\label{ksng-iso}
For any spacetime with $ \bar R = 0 $, with $ \lh $ non-geodesic,
the solution of \ks motions is a subset of its own isometries,
restricted by Eqs.~\rf{eq-sta}. In particular, this holds for any
spacetime of constant curvature.
\epr \N

This proposition seems to be very interesting since it allows an {\em
intrinsic}
reduction of the whole group of isometries of a given spacetime with $ {\bar
R}
=0 $ with the aid of the object $ \lh $, which may, and often does, have a
relevant physical or geometrical content.

From theorem~\ref{isomksng} and proposition~\ref{ksng-iso}, it is
evident that the knowledge of isometries ---for instance, the classical book
\cite{petrov}--- could be applied here in order to develop an extensive
study of
non-geodesic \ks motions. We shall not develop here this fruitful connection
in
detail (see also App.~\ref{app-d}).

To end up this analysis of general properties of non-geodesic \ks motions,
let us add a pair
of  consequences regarding invariant quantities under a non-geodesic \ks
motion.
First, due to theorem~\ref{isomksng} a standard calculation leads to the
conclusion that the Riemannian tensor associated with $ \gamma $ is an
invariant
under a non-geodesic \ks motion. This
constitutes a result
analogous to finding out that the Riemannian tensor or the Weyl tensor of a
metric tensor are the invariant objects under an isometric or a conformal
motion, respectively. Consequently, this allows to interpret non-geodesic
\ks motions as {\em symmetries}, i.e., transformations preserving some
geometrical objects.

Moreover, this result may easily be extrapolated to spacetimes which are
related by a \gks relation, in an analogous way as with the 
well known
invariance of the Weyl tensor under a conformal relation, see e.g.,
\cite{eisen}. The result is
\bpr
\label{inv-ksng}
Let two metric tensors, $ {\bar g} $ and $ g $,
be linked by a \gks relation, that is, $ {\bar g} = g + 2 H \ll $,
with $ \lh $ non-geodesic, then
\beq
  R^{\alpha}_{\; \beta \gamma \delta} [g+
  ( R_{\lm} \ell^{\lambda}  \ell^{\mu}) /a^2)\ll] =
  R^{\alpha}_{\; \beta \gamma \delta} [{\bar g} +
  ({\bar R}_{\lm}\ell^{\lambda}  \ell^{\mu}) /a^2) \ll] . \nnb
\eeq
\epr
Where $ {\bar R}_{\ab} $ is the Ricci tensor of $ {\bar g} $.
We have used the standard convention $ {\bar {\lh}} \equiv \lh $
and also $ {\bar a}^2 = a^2 $, where
${\vec {\bar a}} \equiv {\bar D} \lv $ (see e.g., \cite{Tho,xn2}). \N

A final link between properties of non-geodesic \ks motions and isometries
is worth  mentioning. Since this type of \ks motions have been reduced to a
restricted problem  of isometries, we could also take advantage of constants
of motion along geodesics of $ (V_n, \gamma) $. Obviously, these will not be
in general geodesics of $ (V_n, g ) $. However their study could reveal new
constants of motion for their corresponding
motions in $ (V_n, g ) $ which could shed some light into non-trivial first
integrals in \gks related spacetimes. The kind of motion in $ (V_n, g ) $
which corresponds to the geodesics  of $ \gamma $ will be considered
elsewhere.

To finish, we add an example of non-geodesic \ks motions in $ n = 4 $ which
can be easily visualized geometrically and the solution appears then
natural,
\bex
\label{ex1}
Consider Minkowski spacetime, where $ \{x^\alpha \}$ $ (\alpha=0,1,2,3) $
are
the usual Cartesian coordinates, and $ x^0 $ refers to the timelike
coordinate.
For $ \lh = dx^0 + \alpha \, dx^1 + \beta (\cos \omega x^1 \, dx^2 + \sin
\omega x^1 \, dx^3) $ where $ \beta = \sqrt{1-\alpha^2} $, $ \alpha \in
(-1,1) - \{ 0 \} $ and $ \omega \neq 0 $ are constants, the set of \ks
motions is generated by
$ \{ \partial_{x^0} $, $ \partial_{x^2} $, $ \partial_{x^3} $, $ (1/
\omega) \partial_{x^1} + x^2 \partial_{x^3} - x^3 \partial_{x^2} \} $. In
this case $ m = 0 $.
\eex
\pr Clearly $ \lh $ is null and under the hypothesis $ {\vec a}
= \alpha \beta \omega (-\sin \omega x^1 \, \partial_{x^2} + \cos \omega
x^1\, \partial_{x^3}) \neq 0 $. In this case, from Prop.~\ref{ksng-iso}, \ks
motions are a set of isometries of flat spacetime. The latter are generated
by $ \xiv = A^{\lambda} \, \partial_{x^{\lambda}} + B_i \, (x^i \,
\partial_{x^0} + x^0 \, \partial_{x^i}) + \epsilon_{ij} x^i \,
\partial_{x^j} $ where $ A^{\lambda} $, $ B_i $ $ \epsilon_{ij} = -
\epsilon_{ji} $ are constants. On the other hand, Eqs.~\rf{eq-sta} impose
\beq
\label{eq-mm} m = B_1 \alpha + \beta (B_2 \cos \omega x^1 + B_3 \sin \omega
x^1), \\
m \alpha = B_1 + \beta (\epsilon_{12} \cos \omega x^1 + \epsilon_{13} \sin
\omega x^1), \nnb \\
\label{eq-BB} m \beta \cos \omega x^1 = B_2 - \epsilon_{12} \alpha + \beta
\sin \omega x^1 ( \epsilon_{1i} x^i \omega - A^1\omega - B_1 \omega x^0 +
\epsilon_{23}), \\
m \beta \sin \omega x^1 = B_3 - \epsilon_{13} \alpha - \beta \cos \omega x^1
( \epsilon_{1i} x^i \omega - A^1\omega - B_1 \omega x^0 + \epsilon_{23})
\nnb .
\eeq
From Eq.~\rf{eq-mm} one sees that $ m=m(x^1) $. Therefore from
Eq.~\rf{eq-BB} necessarily $ B_1 =  \epsilon_{1i} = 0 $. Substituting
Eq.~\rf{eq-mm} into Eq.~\rf{eq-BB} one gets $ B_2 = B_3 = 0 $ and $ A^1
\omega = \epsilon_{23} $. The rest of equations become mutually compatible.
\N

Finally, notice that, in general, not for any
$ \lh $ a solution exists, even in flat spacetime.
\subsection{Geodesic $ \ell $ with $ \Delta \ne 0 $}
\label{ss-gldn0}
To begin with, the object $ \Delta $ is a scalar defined by (recall notation
in Sect.~\ref{intro})
\beq
\label{delta1}
\Delta \equiv - 2 D\theta + 4 \theta^2 - 3 {\bar R} + 2 \ell^{\mu}
\nabla_{\sigma} \nabla^{\sigma} \ell_{\mu} + DM - 2M \theta .
\eeq
It can also be written in terms of the Newman-Penrose quantities and the
optical scalars \cite{xn2}
\beq
\label{delta2}
\Delta = 2( \rho^2 + {\bar \rho}^2 - \rho{\bar \rho} - \sigma{\bar \sigma} - 2 \Phi_{00}) = 2(\theta^2 - \varsigma^2- 3 \varpi^2 - {\bar R} + M^2 - 2 M \theta),
\eeq
respectively.
\bpr
\label{prop-par}
$ \Delta $ is only sensitive to changes in the
parametrization of $ \lv $ and in the form $ \Delta_A = A^2\Delta_{A=1} $,
if  $ \lv \rightarrow A\lv $.
\epr
\pr Clearly, $ \Delta $ is an intrinsic scalar associated with $ \lv $. On
the other hand, under the change $ \lv \rightarrow A\lv $ we have
$$\begin{array}{l}
\theta \to A \, \theta + DA/2, \quad \theta^2 \to A^2 \, \theta^2 + A (DA)
\, \theta + (DA)^2/4, \nnb \\
D\theta \to A^2 \, D\theta + A(DA) \, \theta + A(DDA)/2, \quad {\bar R} \to
A^2 \, {\bar R}, \nnb \\
\ell^{\mu} \nabla_{\sigma} \nabla^{\sigma} \ell_{\mu} \to A^2 \ell^{\mu}
\nabla_{\sigma} \nabla^{\sigma} \ell_{\mu}, \quad
M \to A \, M + DA, \nnb \\
DM \to A^2 \, DM + A(DA) \, M + A(DDA). \nnb
\end{array} $$
Then, a direct subsitution proves the assertion. \N

For the sake of brevity, we will only display here the main results of this
case.
We begin with:
\bth
\label{ksgdelta}
The system of geodesic with $ \Delta \ne 0 $  \ks motions is closed.
\eth
\pr The proof is similar to that of theorem~\ref{ksng}. Obviously the
combinations leading to the isolation of $ h $ are different. In our case,
we have found
\begin{eqnarray}
\label{gr} \lie R  = 
2 \{ DDh + (4\theta + M)\, Dh + [2(D\theta + 2 \theta^2)
+ 2 M \theta + DM - {\bar R}]h \}, \hfil{}\\
\ell^{\mu} \lie R_{\mu \alpha} =  [DDh + 2 (\theta + M) \, Dh + (2 D \theta + {\bar R}- \ell^{\lambda} \nabla_{\sigma}
\nabla^{\sigma} \ell_{\lambda} + 2 M \theta
  \nnb \hfil{}\\
\label{gricci} + DM + M^2) h ] \,
\ell_{\alpha}, \\
\label{grie} \ell^{\sigma} \ell^{\mu} \lie R^{\alpha}_{\; \sigma \beta \mu}
=  - [DDh + 3M Dh + 2(M^2 + DM) h ] \, \ell^{\alpha} \ell_{\beta} ,\hfil{}
\end{eqnarray}
where we have made use of some of the expressions given in App.~\ref{app-a}
---recall $ {\vec  a} = M \lv $. Hence we have
\beq
2 \Delta (h \ell_{\alpha} \ell_{\beta}) =
\bigl(\lie R\bigr) \ell_{\alpha}
\ell_{\beta} - 4 \bigl(\lie  R_{\mu (\alpha}\bigr)  \ell_{\beta)}\ell^{\mu}
- 2 \bigl( \lie R_{\alpha \sigma \beta \mu} \bigr) \ell^{\sigma} \ell^{\mu}.
\nnb
\eeq
Clearly if $ \Delta $ is non-zero, $ h $ can be isolated only in terms of
data and the unknowns $ \xi_{\alpha} $, $ \xi_{[\ab]} $. Moreover, notice
that the vanishing of $ \Delta $ is a well-defined condition (see
Prop.~\ref{prop-par}), independent of
the parametrization of $ \lv $.
\N

Again it is possible to rewrite $ h \ll $ in a more compact form, namely,
\bpr
For the case of a geodesic $ \lh $ with $ \Delta \neq 0 $,
\beq
\label{hll-gdeltaks}
2 h \ell_{\alpha} \ell_{\beta}= \lie \Biggl[{R \ell_{\alpha} \ell_{\beta}- 2
\bigl( R_{\alpha \sigma \beta \mu}
\ell^{\sigma} \ell^{\mu} + R_{\alpha \mu} \ell^{\mu} \ell_{\beta} + R_{\beta
\mu} \ell^{\mu} \ell_{\alpha}\bigr)  \over \Delta}\Biggr].
\eeq
\epr
\pr It will suffice to prove $ \lie \Delta = 2 m \Delta $. From
App.~\ref{app-a} ---formulae~i--vi---
we have
\beq
\lie \theta = m \theta + Dm/2 \rightarrow \lie \theta^2 = 2 m \theta^2 +
(Dm)\theta, \nnb \\
\lie D\theta = 2 m \, D \theta + (Dm) \theta + DDm/2, \quad \lie {\bar R} =
2m {\bar R}, \nnb \\
\lie \ell^{\mu} \nabla_{\sigma} \nabla^{\sigma} \ell_{\mu} =
2m \ell^{\mu} \nabla_{\sigma} \nabla^{\sigma} \ell_{\mu}, \quad \lie M = 2m
\, M, \nnb \\ \lie DM = 2m \, DM + (Dm)M + (DDm) \nnb.
\eeq
Consequently, $ \lie \Delta = 2m \Delta $. Now one can follow similar
steps as in the proof of theorem~\ref{isomksng} in order to get
expression~\rf{hll-gdeltaks}. \N

Expression~\rf{hll-gdeltaks} allows us to rewrite Eq.~\rf{eq-h} as
\beq
\lie \gamma = 0, \ {\rm with} \ \gamma_{\ab}  \equiv g_{\ab} + {2 \bigl(
R_{\alpha \sigma
\beta \mu} \ell^{\sigma} \ell^{\mu} + R_{\alpha \mu} \ell^{\mu} \ell_{\beta}
+ R_{\beta \mu} \ell^{\mu} \ell_{\alpha}\bigr) -
R \ell_{\alpha} \ell_{\beta}
\over \Delta}. \nnb
\eeq
Despite the non-geodesic case, $ \gamma $ is {\em not always} linked by a
\gks relation with $ g $. Moreover $ \gamma $ may become degenerate
---although it can only be completely degenerate for $ n=3 $, as is easily
seen from a study of $ \gamma_{\alpha \lambda} \ell^{\lambda} $ and the
trace of $ \gamma $. Its degeneracy in $ n= 4 $ is controlled by the
following result:
\bpr
\label{deg-of-delta}
For $ n = 4 $, the determinant of $ \gamma $, in any orthonormal cobasis,
equals to $ (1/ \Delta^4) \allowbreak\{(\Delta+
2 {\bar R})^2\allowbreak\bigl[4\allowbreak ||\Psi_0 ||^2-(\Delta + {\bar
R})^2 \bigr]\} $.
\epr
\pr Using the Newman-Penrose formalism ---see e.g., \cite{kramer,chandra}
for
definitions of each object--- we write
$$\begin{array}{rcl}
\gamma_{\ab} & = & g_{\ab} + \left( {2 \over \Delta} \right) [(\Psi_2+ {\bar
{\Psi}}_2 - 2 \Lambda - 2 \Phi_{11}) \ell_{\alpha} \ell_{\beta} + ({\bar
\Phi}_{01} - {\bar \Psi}_1)(\ell_{\alpha} k_{\beta} + k_{\alpha}
\ell_{\beta}) \nnb \\
& & + (\Phi_{01} - \Psi_1)
(\ell_{\alpha} {\bar k}_{\beta} + {\bar k}_{\alpha} \ell_{\beta})
+ {\bar \Psi}_0 k_{\alpha} k_{\beta} + \Psi_0 {\bar k}_{\alpha} {\bar
k}_{\beta} \cr 
&&+ \Phi_{00} (g_{\ab} - m_{\alpha} \ell_{\beta} - m_{\beta}
\ell_{\alpha})],
\end{array} $$
where $ \{ \lh, \mh, \kh, \bkh \} $ is
a null cobasis containing $ \lh $.
Indeed, any term proportional to $ \ll $ will not affect the value of the
determinant of $ \gamma $. Therefore, we only need to consider $ {\tilde
\gamma} \equiv \gamma - (2/\Delta) ( \Psi_2 + {\bar \Psi}_2 - 2 \Lambda - 2
\Phi_{11}) \ll $.

In terms of an orthonormal cobasis, $ \{\th^0, \th^1,\th^2,\th^3 \} $,
related with the previous one by $ \lh = (\th^0 + \th^1)/\sqrt{2} $, $ \mh =
(\th^0 - \th^1)/\sqrt{2} $, $ \kh = (\th^2 + i \th^3)/\sqrt{2} $, we get
$$\begin{array}{rcl}
{\tilde \gamma}
&=& \biggl( 1 + {4 \Phi_{00} \over \Delta} \biggr) ( - \th^0
\otimes \th^0 + \th^1 \otimes \th^1 ) + 
\biggl( 1 + { 2 \Phi_{00} + \Psi_0 +
{\bar \Psi}_0 \over \Delta} \biggr) \th^2 \otimes \th^2 \cr
& &+ \biggl( 1 + { 2
\Phi_{00} - \Psi_0 - {\bar \Psi}_0 \over \Delta} \biggr) \th^3 \otimes \th^3 \cr
& & + {(\Phi_{01} + {\bar \Phi}_{01} - \Psi_1 - {\bar \Psi}_1) \over
\Delta} (\th^0\otimes\th^2 + \th^2 \otimes \th^0 + \th^1 \otimes \th^2 +
\th^2 \otimes \th^1) \cr
& & + {i({\bar \Phi}_{01} - {\Phi}_{01} + \Psi_1 - {\bar
\Psi}_1) \over \Delta} (\th^0\otimes\th^3 + \th^3 \otimes \th^0 + \th^1
\otimes \th^3 + \th^3 \otimes \th^1) \cr
& & + {i({\bar \Psi}_0 - \Psi_0) \over \Delta}
(\th^2\otimes\th^3 + \th^3\otimes\th^2) .
\end{array} $$
Then a standard computation gives
$$ \det {\tilde \gamma} = (1/ \Delta^4) \allowbreak \{(\Delta+
4 \Phi_{00})^2\allowbreak\bigl[4\allowbreak ||\Psi_0 ||^2-(\Delta + 2
\Phi_{00})^2 \bigr]\} .$$
Recalling that $ 2 \Phi_{00} \equiv {\bar R} $ and $\det {\tilde \gamma} =
\det \gamma $ we get the result claimed above. \N

It is worth remarking, as an example, that $ \gamma $ is a metric if $ g $
is the metric of
any spacetime of constant curvature. This happens because, for
a spacetime of constant curvature, one has
$ R_{\alpha \beta \gamma \delta} = k (g_{\alpha \gamma} g_{\beta \delta}
- g_{\alpha \delta} g_{\beta \gamma}) $, where $ k $ is a constant related
with the (constant) scalar curvature by: $ k = R/n(n-1) $. Whence,
$ {\bar R} = R_{\lm} \ell^{\lambda} \ell^{\mu} = (R/n) g_{\lm}
\ell^{\lambda} \ell^{\mu} = 0 $. Moreover, a spacetime of
constant curvature is conformally flat. Therefore $ \Psi_0 = 0 $ and
$ \det \gamma = -1 $, in any orthonormal cobasis.
Although $ \gamma $ will be {\em in general non-degenerate},
we do not have now a direct copy of
theorem~\ref{isomksng}, but a  similar result,
\bpr
[Isometrization of geodesic $ \Delta \ne 0 $ \ks motions]
\label{isomksg-delta}
$\!\!$The $ \! $KS\-VFs of a geodesic
\ks motion with $ \Delta \neq 0 $ are
vector fields associated with the
invariance of the object $ \gamma $,
restricted by Eqs.~\rf{eq-sta}.
Moreover, when $ \det \!\gamma \ne 0 $, the vector fields
associated with the invariance of $ \gamma $ are Killing vector fields.
\epr \N

Notice that $ \gamma $ is 
not always linked with $ g $ by a \gks relation, and, therefore, $ \lie
\gamma = 0
$ does not imply by itself Eq.~\rf{eq-h} in general.

We also have a similar result as in Prop.~\ref{ksng-iso}.
First define $ T_{\ab} =  \bigl[2 ( R_{\alpha \sigma \beta \mu} \ell^{\sigma}
\ell^{\mu}
+
R_{\alpha \mu} \ell^{\mu} \ell_{\beta} + R_{\beta \mu} \ell^{\mu}
\ell_{\alpha})
- R \ell_{\alpha} \ell_{\beta}\bigr] / \Delta $. Then
\bpr
\label{ksg-delta-iso}
For any  geodesic $ \lh $ with $ \Delta \ne 0 $ and in any spacetime with $
T = 0 $, the solution of \ks motions is a subset of its own isometries,
restricted by Eqs.~\rf{eq-sta}.
\epr \N

Similar considerations as those in previous section are valid now, e.g.,
Prop.~\ref{inv-ksng}, for the case of {\em non}-degenerate $ \gamma $, and
its
constants of motion. They also justify to interpret this case of \ks motions
as
some kind of {\em symmetries}. These results are now more interesting
because $ \lh $ is clearly related with the light-cone structure of a
spacetime. Inside this case most of the more used $ \lh $ in General
Relativity are to be found, e.g., the axially symmetric case or the
spherically symmetric case. These examples are carried out in the following
section.
\subsubsection{KSVFs for the principal null directions of Kerr-Newman
spacetimes}
\label{sss-kskn}
In this section we shall focus on the resolution of Kerr-Schild motions for
a class of spacetimes which are of major astrophysical interest, i.e.,
Kerr-Newman spacetimes.

\proclaim Formulation of the problem.

\ni The equations to be solved are Kerr-Schild
equations where $ \bg $ is
now the metric of Kerr-Newman
spacetimes and $ \lh $ is any of their two principal null directions ---see
below.

One way to solve the equations is to consider them
directly, i.e., expanding them in terms of a partial derivative system.
Yet this way is long. Indeed, there is an alternative path that makes use
of some previous results about Kerr-Schild motions. Moreover, it may prove
to be useful in other spacetimes, too.

The calculations are simplified by noticing that
Kerr-Newman metrics are Kerr-Schild metrics, i.e.,
\beq
\label{eq-knk}
\bg_{KN} = \eta + 2 H \ll,
\eeq
where $ \eta $ is the metric tensor of flat spacetime, and in terms
of  ``Cartesian-like'', or ``Kerr-Schild'' coordinates, $ H $ and
$ \lh $ are expressed as \cite{kramer,mtw,chandra}:
\beq
\label{eq-H}
H(x,y,z) = {2M r - Q^2 \over r^2 + a^2 (z/r)^2},\quad
r^2(x^2+y^2+z^2) + a^2 z^2 = r^2(r^2 + a^2),
\eeq
where $ M $ and $ Q $ represent the mass and the charge of the
source,~$ a $ is its angular momentum per unit mass, and
\beq
\label{eq-l}
\lh_{(\pm)} = {1 \over \sqrt{2}} \Biggl(\pm\, dt + {z \over r} \, dz +
{rx + ay \over r^2 + a^2} \, dx + {ry - ax \over r^2 + a^2} \, dy \Biggl)
\eeq
are the two principal null directions of Kerr-Newman spacetimes, which
are geodesic and shear-free.

First, Prop.~\ref{prop-two} assures that two metrics related by a \gks
relation,
$ {\btg} = \bg + 2H \ll $, admit the {\em same} KSVFs with respect to $ \lh
$.
The relation between the functions $\bar h $, $ \bar m $,
$ h $ and $ m $ is, in this case: $ \bar h = h + \lie H + 2m H $,
$ \bar m  = m $.

Thus, the problem
\beq
\lie \bg_{KN} = 2 h_{KN} \ll , \quad \lie \lh =m \lh \nnb
\eeq
is equivalent to
\beq
\label{eq-f}
\lie \eta = 2 h_F \, \ll , \quad \lie \lh = m \lh,
\eeq
where ``$ F $'' refers to flat spacetime and $ \xiv $ are
the {\em same} set in both cases.
We have also proven that the
solution to a problem of Kerr-Schild
motions for a geodesic $ \lh $ depends on whether
$ \Delta $ ---Eqs.~\rf{delta1},~\rf{delta2}--- vanishes or not.

Thanks to Prop.~\ref{prop-two}, the calculation is simplified
to the computation of $ \Delta $ for~\rf{eq-l}
where $ \{ x,y,z, t \} $ are now Cartesian coordinates and where
computations are much easier.
Moreover, we can make use of some general results
concerning the vanishing of $ \Delta $ in flat spacetime.
In Sect.~\ref{sss-sld0} this issue will be solved when $ \lh $ is a
principal null direction and $ {\bar R} = 0 $ (Clearly this includes our
case).
The result is that for a shear-free $ \lh $, $ \Delta = 0 $ if in addition
$ \lh $ is rotation and expansion-free. However, $ \lh_{\pm} $ possesses
rotation unless $ a = 0 $. And if $ a= 0 $, $ \lh $ is simply
given by
\beq
\lh = {1 \over \sqrt{2}} \bigl(\pm \, dt + dr \bigr), \quad
r^2 = x^2 + y^2 + z^2 , \nnb
\eeq
which satisfies $ \Delta = 1 / r^2 \neq 0 $, as is easily computed.
Summarizing, we get\footnote{Indeed it is not difficult to show that $
\Delta $ does not vanish as well in Kerr-Newman spacetimes. The reason is
that $ \Delta $ remains {\em invariant} if there exists a Kerr-Schild
relation between both spacetimes in which $ \lh $ is geodesic.
This is accomplished in Eq.~\rf{eq-knk}.}
\bpr
For $ \lh $ given in Eq.~\rf{eq-l}, $ \Delta $ does not vanish in
flat spacetime.
\epr

\proclaim Resolution of \ks motions.
\newline

With this result in hand,
we use Prop.~\ref{isomksg-delta} in order to conclude
that Eqs.~\rf{eq-f} reduce to
\beq
\lie \eta = 0, \quad \lie \lh = m \lh . \nnb
\eeq
The solution of the first set is simply the generators of the Poincar\'e
group. In order to solve the second set, one only needs to consider a
generic infinitesimal generator of the Poincar\'e group, Eq.~\rf{eq-l}
for $ \lh $  and impose
\beq
\label{eq-bra}
[\xiv, \lv ] = m \lv .
\eeq

We choose the following representation
of a general infinitesimal generator of the Poincar\'e group which is
clearly adapted to our purposes:
\beq
\label{eq-gis}
\begin{array}{rcl} \xiv & = & (\alpha_1 + \beta_1 x + \beta_2 y + \beta_3 z)
\partial_t + (\alpha_2 + \beta_1 t - \gamma_1 y - \gamma_2 z ) \partial_x
\cr
& & + (\alpha_3 + \beta_2 t + \gamma_1 x - \gamma_3 z ) \partial_y +
(\alpha_4 +
\beta_3 t + \gamma_2 x + \gamma_3 y ) \partial_z,
\end{array}
\eeq
where $ \alpha_{\lambda} $, $ \beta_i $, $ \gamma_i $ are constants. Taking
into account~\rf{eq-l},~\rf{eq-gis} and~\rf{eq-bra}, one
gets that the terms with $ \partial_t $ yield:
$ m = \beta_1 \ell_x + \beta_2 \ell_y + \beta_3 \ell_z $.
However, combining the latter result with the terms with
$ \partial_x $, one obtains, $ m = 0 $ (and therefore $ \beta_i = 0 $)
and $ \alpha_2 = 0 $. Then, from some of the terms with
$ \partial_y $, $ \partial_z $, one easily gets $ \alpha_3 = \alpha_4 = 0 $.
Finally, the remaining conditions for the $ \partial_x $ terms are
\beq
{a^2(a^2x-xr^2-2ary)zr \over (r^2 + a^2)^2(r^4 + a^2 z^2)}
(\gamma_2 x + \gamma_3 y) - {a z \over r^2 + a^2} \gamma_3
+ {a^2z \over r(r^2 +a^2)} \gamma_2 = 0, \nnb
\eeq
where we have used $ \partial_x r = xr^3/(r^4+a^2z^2) $,
$ \partial_x r = yr^3/(r^4+a^2z^2) $ and $ \partial_z r =
zr(r^2 + a^2)/(r^4 + a^2 z^2) $. There are clearly two
situations: $ a = 0 $ and $ a \neq  0 $. In the first situation
all $ \gamma_i $ remain free. The same holds for the
terms with $ \partial_y $ and $  \partial_z $ as one can readily
check. The result is therefore $ \{ \xiv \} = \{ \partial_t, x\partial_y - y
\partial_x,
x \partial_z - z \partial_x, y \partial_z - z \partial_y \} $ which
corresponds to the (irrotational) radial case or spherically symmetric case.
The other situation is $ a \neq 0 $. In this case, $ \gamma_2 $
and $ \gamma_3 $ must be zero necessarily. The only remaining
parameter is $ \gamma_1 $. One then calculates the remaining terms
with $ \partial_y $ and $ \partial_z $. The $ \partial_y $ terms yield
the same conditions as the $ \partial_x $ ones, as expected. Finally
the $ \partial_z $ terms are identically satisfied because
$ (x\partial_y - y\partial_x) r =0 $. Thus $ \alpha_1 $ and $ \gamma_1 $
remain free. This yields
$ \{ \xiv \} = \{ \partial_t, x \partial_y - y \partial_x \} $ for the
rotational case.

Let us recall that the infinitesimal generators have multiple
representations depending on the
coordinate system being used. For the \ks one, the result is the one
displayed before. In Kerr coordinates
the result is $ \{ \xiv \} = \{ \partial_t, \partial_{\phi} \} $ for the
rotational case, for instance, where the relation
between both system of coordinates is given by
\beq
\label{eq-rcb}
x + i  y = ( r + i a ) e^{i\phi} \sin \theta, \quad z = r \cos \theta. \nnb
\eeq

So far, this is the solution for flat spacetime. We have shown before that
the infinitesimal
generators are the same for  Kerr-Newman spacetimes. Yet we can
easily find their action on Kerr-Newman metrics.
In our case, $ m = 0 $, $ h_F =0 $, and we have $ H = H(r,z) $ for $ a \neq
0 $, and
$ H = H(r) $ for $ a= 0 $. The result is, in any case $  h_{KN} = m  = 0 $.
Therefore, the \ks motions are the isometries of Kerr-Newman spaces.
Summarizing
($ T_{x^{\alpha}} $ stands for translation along the
axis $ x^{\alpha} $)
\bpr
\label{npkskn}
There are no proper KSVFs for Kerr-Newman spacetimes and their
principal null directions. \N
\epr
\bpr
\label{prop-kskn}
The \ks motions for KN spaces associated with the principal
null directions are given by $ T_{t} \otimes T_{\phi} $ for rotational
$ \lh $, and by $ T_{t} \otimes SO(3) $ for the irrotational case. \N
\epr
Here $ t $ and $ \phi $ are Kerr coordinates, not Boyer-Lindquist
coordinates. In \ks coordinates, cf. Eqs. above, $ T_{\phi} $ is
equivalent to $ R_z $, i.e., a rotation around the $ z $-axis. Another
consequence is: \footnote{Of course, we could consider any $ H $, yet we
focus on spacetimes with the {\em same} local motions, and therefore they should
share a similar action upon the metric tensor.}
\bcr
For any \ks metric of the form $ \bg = \eta + 2 H \ll $,
where $ \lh $ are the null directions given in~\rf{eq-l}, $ H = H (r, z/r) $
for $ a \neq 0 $, and $ H = H(r) $ for $ a = 0 $,  the solution to the
problem of \ks motions is given by Prop.~\ref{prop-kskn}. In particular,
this also includes flat spacetime. \N
\ecr
Indeed, this result points to an idea on physical applications of \ks
symmetries
which is considered elsewhere \cite{pakss}. Let us mention that among the
spacetimes of
this family one finds {\em all} proposed candidates to describe the
macroscopic properties of non-singular quantum interiors for black holes,
with a clear
source origin, which are currently under study,
see e.g., \cite{behm}.

Another remarkable example of this section are
\ks motions for $ n
$-dimensional flat spacetime and a spherically symmetric deformation
direction.
\bex
\label{spherical}
In flat spacetime, the \ks motions for $ \lh = d(t \pm r) $, where $ r $ is
the usual radial coordinate defining the radius of the $ (n-2) $ dimensional
spheres, are $ SO(n-2) \times T_{x^0} $, where $ x^0 $ is the timelike
coordinate. Moreover, $ m = 0 $.
\eex
\pr This result is proven in I for $ n= 4 $; analogous steps as those prove
the latter result. \N

We remark that in spite of
the fact that for all these spacetimes the KSVFs
correspond to Killing vector fields, one should not conclude that
\ks motions seem to be a simple subset of isometries. In
I several examples with {\em proper} \ks motions are solved. And even
for the cases where \ks motions reduce to isometries, they may be important
because they are a subset of isometries restricted by a condition which is
related with the invariance of the congruence of curves associated with $
\lh $ and, therefore, with {\em the light-cone structure of spacetime}
itself ---see also Exs.~\ref{exlm} later on.
\subsection{Geodesic $ \ell $ with $ \Delta =  0 $}
\label{s-gld0}

The remaining case, i.e., $ \lh $ which are geodesic but satisfy
$ \Delta = 0 $, is still without a complete solution. There are however
some points
worth to be remarked. First, the existence of such null vectors is severely
restricted in general, except in flat spacetime, where the maximum number of
such vectors is obtained. Indeed, the condition is very similar to that of
the well known Goldberg-Sachs theorem and its generalizations (see e.g.,
\cite{kramer}). In principle one can use the Newman-Penrose formalism to
perform a complete study. In Sect.~\ref{sss-sld0} we have begun with this
idea for  $\lh $ which are principal null directions and satisfy $ \bar R =
0 $.

Secondly, one should remember that within this subcase, the system remains
{\em open}. This tells us that {\em no} general solution similar to the
previous cases may be attained now for all $ \lh $. In fact, one needs to
classify, explicitly, the $ \lh $'s which make the system open and only a
particular
study will reveal the characteristics of their infinite dimensional
algebras. For the rest, one is faced with a closed problem. Since the
existence of geodesic $ \lh $ with $
\Delta = 0 $ completely depends on each spacetime, we do not have the
confidence that a global result will be found, but rather a detailed
collection of solutions. The unique exception is the
two-dimensional problem. For $ n= 2 $ any $ \lh $ belongs to this case and
the general solution, containing infinite dimensional algebras, was given in
I. This result is completely similar to that of conformal symmetries.

Notice, however,  that the geometrical object $ G $ defined by $ G\!
\equiv\!
R_{\alpha
\sigma \beta \mu} \ell^{\sigma} \ell^{\mu} + R_{\alpha \mu} \ell^{\mu}
\ell_{\beta} + R_{\beta \mu} \ell^{\mu} \ell_{\alpha} - (R/2) \ell_{\alpha}
\ell_{\beta} $ remains conformally invariant under a geodes\-ic, $\Delta = 0$
\ks motion, i.e., $ \lie G = 2m G $. This is obtained from~\rf{eq-sta}
and~\rf{hll-gdeltaks}. This show us that \ks motions act in this case as a
{\em symmetry} in some sense as well.

Among several possible examples 
we refer the reader to
two examples given in I which show all the special characteristics of
this case. These are the \ks motions in an $ n $-dimensional 
flat spacetime for an $ \lh $
adapted to the cylindrical symmetry and to the parallel symmetry. Both have very peculiar features in their Lie algebras. 
For instance, in the parallel case the Lie algebra is of an infinite dimensional character. Although we will not discuss them here in detail, Ex.~\ref{exlm} ---Sect.~\ref{ss-lglm}---
shows the solution of the parallel case 
in the four dimensional case. 
\subsubsection{Subclassification of $ \ell $ with $ \Delta = 0 $}
\label{sss-sld0}
In this section we would like to analyze further the
classification of
a geodesic $ \lh $ that satisfies $ \Delta = 0 $ for the physical
interesting case of Minkowski space-time ---where a maximum of such vectors
may exist. The results are,
however, easily generalized to other spacetimes as will appropriately be
pointed out in the text.

The study of the existence of such $ \lh $ in any spacetime is
better carried out with the Newman-Penrose formalism \cite{kramer}. In our
case $ \bar R (= 2 \Phi_{00}) $ cancels. Furthermore, without losing
generality, we shall
take $ M $ equal to zero (recall Prop.~\ref{prop-par} $ \Delta  = 0 $ is
invariant under changes
in the parametrization of $ \lh $). Then we have from~\rf{delta2}
\beq
\label{eq-main}
\rho^2 + {\bar \rho}^2 = \rho {\bar \rho} + \sigma {\bar \sigma}.
\eeq

The Newman-Penrose equations that are of relevance for our equation are
\beq
\label{eq-rho}D \rho = \rho^2 + \sigma \bsigma, \quad D \sigma = \sigma(\rho
+
\brho + 4 \epsilon) .
\eeq
Deriving~\rf{eq-main} with respect to $ \lv $ and using~\rf{eq-rho}, we get
\beq
\label{eq-rho-sigma}
2\rho^3 + 2 \brho^3 = (\rho + \brho)(\rho\brho + \sigma\bsigma).
\eeq
Using~\rf{eq-main} and~\rf{eq-rho-sigma} we obtain
\beq
(\rho + \brho)(\rho\brho-\sigma\bsigma) = 0 . \nnb
\eeq
Two possibilities appear. The first one is when $ \rho = - \brho $.
Substituting
this condition into Eq.~\rf{eq-main}, we obtain $ 3 \rho^2 = \sigma \bsigma
$.
Taking again the derivative with respect to $ \lv $, we get that this
condition
is only fulfilled if $ \rho = \sigma = 0 $. In terms of the optical scalars,
this means that the affine parametrized geodesic null vector must satisfy $
\theta = \varsigma = \varpi = 0 $.

The other possibility is $ \rho \brho = \sigma \bsigma $. Substituting it
into
Eq.~\rf{eq-main}, we get the condition $ \rho = \brho $ ($ \varpi = 0 $) and
therefore $ \sigma \bsigma = \rho^2 $. Provided that $ \lh $ is geodesic and
$
\Phi_{00} $, $ \Psi_0 $ are zero, one can apply the Sachs theorem,
\cite{sachs},
to obtain that all possible such $ \lh $ must belong to one of the following
sets
\beq
\cases{\rho =- \sigma = -{1 \over 2 s} & if $  \rho \neq 0 $, \cr
\rho = \sigma = 0 & if $ \rho = 0 $, \cr} \nnb
\eeq
where $ s $ is a convenient affine parameter along the congruence generated
by $ \lv $.
We remark that these conclusions are
also valid for any principal null direction of a spacetime for which $ \bar
R = 0 $ (in terms of Newman-Penrose quantities:
$ \Psi_{0} = 0 $ and $ \Phi_{00} (=\bar R/2) = 0 $).
Finally, let us recall that \gks relations used in physics have almost
always
$ \lh $ as a principal null direction.

Eventually, in Minkowski spacetime and with the aid of an extension of the
Kerr theorem,
\cite{kramer}, it
is possible to demonstrate that the conditions $ \rho = \sigma =0 $, $ k = 0
$,
$ \epsilon + \bar\epsilon = 0 $ imply that $ \lh $ is
covariantly constant.
Whereas $ \rho = -\sigma $, $ \bar \rho = \rho  $, or
equivalently $ \varpi = 0$, $ \varsigma^2 = 4 \theta^2 $,
seems to force $ \lh $ to be the cylindrical case
(see e.g., I and Ex.~\ref{exlm}
for their explicit solutions). In that case the study of \ks motions for
geodesic $ \lh $ with $ \Delta = 0 $ in Minkowski spacetime would be
complete.
\subsection{Summary on \ks motions}
\label{ss-sksm}
In previous sections,
we have investigated the possibility of solving the system of
differential equations of other types
of metric motions
than the ones of isometric or conformal motions.
Specifically, we have studied in detail
the existence of \ks motions.
In particular, we have
shown how the problem is divided into
three main different cases. In two of them, whenever $ \lh $
is non-geodesic {\em or} is actually
geodesic but satisfies $ \Delta \neq 0 $ (Sects.~\ref{ss-ngk}
and~\ref{ss-gldn0}), the problem of \ks motions yields a finite
dimensional algebra.
For each case several
properties and some examples have been
presented and discussed. Among them, we
have obtained the
main set of geometrical invariants for these cases.
This result is important, and allows one to
consider \ks motions as \ks {\em symmetries}.

Moreover, we  have
shown that \ks symmetries can be linked with
{\em a restricted problem of isometries} of another
Riemannian manifold, except for very peculiar cases. This constitutes
an extension of Refs.~\cite{yano}, \cite{bilyalov}--\cite{hst} for \ks
motions. The reduction is
far from being trivial and adds a further ---geometrical---
path to group isometries
(see examples and App.~\ref{app-d}) ---in an intrinsic way that may
be helpful in the search of physical applications. Finally it
gives a simple and direct procedure for solving the problem of the
symmetry in some relevant cases.

In Sect.~\ref{sss-kskn}, we have given the solution of \ks motions
for spacetimes of great astrophysical interest, Kerr-Newman
spacetimes, pointing to
some physical applications.

The third case, Sect.~\ref{s-gld0}, only exists under
very restrictive conditions. Nevertheless, the spacetimes
and null vectors satisfying such conditions are of relevance,
as is the case of an $ n $-dimensional flat spacetime with
cylindrical or parallel direction.
It is actually the most difficult one.
There, the system remains {\em open}, i.e., without any further and
external condition, the freedom in the unknowns is functional.
Thus, infinite dimensional Lie algebras appear. This is the
{\em first} time that in a metric symmetry such behavior shows up.
Those facts are rather representative of what might be expected  from the
whole family of metric symmetries. However, everything has to be understood
under a different scheme than the one used for
isometric and conformal motions. Such general scheme
will be dealt with in the following sections).

Finally, Sect.~\ref{sss-sld0} concluded focussing on \ks
motions in spacetimes where $ \lh $ is a principal null direction and $
{\bar R} = 0 $. In particular, in flat spacetime it was seen that the
knowledge of \ks motions is almost completed. A
necessary result to finish the study of the more
elaborate situation of \ks motions. In App.~\ref{app-a}, we give some
formulae for the Lie derivatives with respect a KSVF of some basic geometric
objects used along previous sections.
\section{Metric motions. A geometrical approach}

The aim in
previous sections was not focussed on a deeper study
of isometric or conformal groups, nor in possible new applications of both.
Rather, the scope was to try to enlarge the number of metric
motions by
considering a new specific example, \ks
motions. The aim of the following
sections is to pose and develop a framework to deal with some general
situations, since one expects that many achievements of differential
geometry on Riemannian spaces have a physical spin-off. Moreover, \ks
symmetries have some general relevant features that have impelled
us, along with other researchers, to consider metric motions from a
more general point of view, mainly focussed on analyzing the whole set of
candidates to become metric motions, which have a geometrical origin. We
have decided to call them ``generalized'' metric motions, or, simply, metric
motions when no doubt can arise.

In Sects.~\ref{s-gmm},~\ref{ss-dgmg} and~\ref{ss-ivmg} we elaborate this
last idea. In Sect.~\ref{s-mmgd},
we will consider new candidates
of metric groups within the proposed framework, including detailed examples
in some relevant situations.

For other works that introduce some particular generalized metric motions
see, for instance, the works of N.H. Ibragimov
\cite{ibragimov,ibragimov2}, the ones of
N. Muppinaiya \cite{muppi}, the ones of B.C. Xanthopoulos and K.E.
Mastronikola  \cite{xanta1}--\cite{masxan}, and the ones of
Ll. Bel \cite{bel98}. All of them can easily be included in the following
framework.

\subsection{Generalized metric motions}
\label{s-gmm}

Our scope in this section is to formulate
a proper definition of a  generalization of
isometries and conformal symmetries within a geometrical  framework.
Thus, we will first introduce some motivations and, eventually, will
present our choice in definitions~\ref{def-ms} and~\ref{def-ms2}
---Sects.~\ref{ss-dgmg} and~\ref{ss-ivmg}.
\subsubsection{A remark concerning possible physical applications}
\label{ss-rcpa}
An important issue in General Relativity is to find new solutions to
Einstein's field equations with a clear physical use. There are several ways
to achieve that aim. An outstanding example are, e.g.,
\ks relations~\cite{ks} (see also I for a brief review).

Few years after the introduction of the \ks relation, J.F. Pleba\'{n}\-ski and
A. Schild introduced a generalization. Their choice was~\cite{ps}
$$ {\bar g} = g + 2F \mm + G (\slm) + 2H \ll, $$
where $ {\bar g} $ is a rank-two symmetric
tensor (in the case of having a
Lorentzian signature it is to be identified with a metric tensor of a
spacetime), $ g $ is a given metric tensor, $ \lh $, $ \mh $ are two given
null 1-forms and $ F $, $ G $, $H $ are some functions. Its study has given
some interesting results as e.g., regarding Kerr-Newman solutions and
complex
relativity (see also \cite{hhmci}). However, a detailed analysis still
lacks, due in part to the
complexity in the computations. Clearly, the \gks relation is
obtained setting either $ F = G = 0 $ or $ G = H = 0 $.

On the other hand, in the early 90s, S. Bonanos~\cite{bonanos} introduced
another generalization
of the \ks relation. His choice was
$$ {\bar g} = g - (\spq) + p^2 \, \qq, $$
where $ {\bar g} $ may always be interpreted as a metric tensor
---the signature of $ \bar g $ is that of $ g $---,
$ g $ is a given
metric tensor, and $ \ph $ and $ \qh $ are two orthogonal 1-forms which may
be spacelike, timelike or null. In that work some well known spacetimes
were recovered and the formalism was
adapted for studying vacuum
solutions. However, computations are again difficult
and its development still remains.

In these two examples, as well as in any further attempt, it is very
important to
have some knowledge on the wealth of new results. For that reason, the
knowledge of the cases that give rise to an equivalent metric tensor are
fundamental in order to avoid them and centre the computations in new
solutions.  We note that this can be posed as a
problem of {\em internal} transformations, i.e., transformations of a metric
tensor into itself. In this sense, some metric motions ---to be defined
later--- become a useful tool, if not the natural one to ascertain which metric 
relations are redundant. Indeed, \ks motions are a
first example. That is, the knowledge of \ks motions of a given metric
tensor informs us on those \ks relations which do not yield in physical terms a new metric tensor. Let us now present other motivations for defining new metric
motions.
\subsubsection{Preliminary features on known and generalized metric motions}
\label{ss-pfgm}
To begin, let us recall the two well known cases of metric
motions, i.e., isometries and conformal ones. In both situations,
using standard notation, one writes, respectively,
\beq
\label{eq1}
\rea (\bg) = \bg , \quad \rea(\bg) = e^{2 \Phi_t} \bg ,
\eeq
where $ \re $ stands for the pull-back application associated
with the  diffeomorphism $ \varphi ( x ) $, $ t \in (-a,a) $ for some appropriate $ a \in {\bf R} $, $ \bg $ is the  metric tensor and $
\exp 2 \Phi_t $ is the conformal factor.

Besides these two situations, it is not difficult
to imagine other kind of actions over $ \bg $ ---e.g., \ks motions.
Among possible generalizations, we are
interested in those cases which admit a
{\em geometrical interpretation}, i.e.,
{\em those which are generated by the appearance
of other geometrical objects besides $ \bg $
in the r.h.s. of Eqs.~\rf{eq1}}. Although this
is not the only possibility, it follows a viewpoint which
is close to the main line of thought used in the studies of
other types of motions ---e.g., in collineations.
Moreover, it is an interesting choice, since
it often happens that some geometrical structures have a direct connection
with physical issues ---recall the \ks case and previous section.

Even though starting with \ks motions would suffice to introduce
the main features for the definition of generalized metric motions, we will
present a slightly different situation which will help in noticing that a
general
framework for metric motions is indeed possible and that it is similar to
that of the \ks case. A possible form of a new action could be, for
instance,
\beq
\rea (\bg) = \bg + {\Psi}_t \,\uh \otimes \uh, \quad \hbox{or} \quad \rea
(\bg) = e^{2\Phi_t}\bg + {\Psi}_t \, \uh \otimes \uh , \nnb
\eeq
being $ \uh $ a 1-form field of the manifold. To fix ideas only,
but without losing  generality, let us consider that the manifold has a
Lorentzian signature. Then $ \uh $ could be, for instance, a normalized
timelike
vector representing the four-velocity field of a fluid (obviously,
in the case of a null 1-form field, one would recover
the case of \ks motions).
This new proposal is neither an isometry nor a conformal motion nor a \ks motion. Therefore,
it can be considered
as a new problem of metric motions. Because this is an introduction, and for
the sake of simplicity, we will just focus on the non-conformal attempt.
Let us call it the ``{\sl tt}-like'' motion in the sequel.

It is known that in any {\em local} 1-parameter group of local diffeomorphisms, it is
to be satisfied that, whenever $ \varphi_t $, $ \varphi_{t'} $, $ \varphi_{t+t'} $ are well defined,
\beq
\re_{t'} \circ \rea = \re_{t + t'}. \nnb
\eeq
Here $ \circ $ stands for the composition of applications. This property,
once translated into our problem, is written as
\beq
\label{compo}
\re_{t'}(\rea(\bg)) = \re_{t+ t'}(\bg) .
\eeq
Thus, one should impose in our example, that
\beq
\label{eq2}
\re_{t'}( \bg + {\Psi}_t\,\uh \otimes \uh ) =  \bg + \Psi_{t+ t'} \uu .
\eeq
One may recall this basic
property as the {\em property of stability}
of the considered metric motion, since it ensures that a given metric
transformation will belong to the same type regardless
of the times it is applied.
In the case of isometric and conformal motions, one readily
recovers the usual result for them, namely, that there are no
further consequences coming from Cond.~\rf{compo}.

Continuing with Eqs.~\rf{eq2}, we write
$$
\bg + \Psi_{t'} \uu + \Psi_t (\varphi_{t'}(x))
\bigl[\re_{t'}(\uh)  \otimes
\re_{t'}(\uh) \bigr] = \bg + \Psi_{t+t'} \uu. $$
Whence, one deduces
a necessary, and sufficient, law of transformation
for $ \uh $, namely
\beq
\label{eq3}
\re_t (\uh ) = B_t \, \uh ,
\eeq
being $ B_t $ a $ C^{\infty} $ function of $ t $. This is in fact analogous
to the case of \ks motions,

Thus, the {\sl tt}-like motion in its initial form, i.e.,
\beq
\label{eqtt}
\rea (\bg) = \bg + \Psi_t \,\uu
\eeq
implies necessarily Eq.~\rf{eq3}.
Then, $ B_t $ may be re-interpreted as a new unknown of the complete
problem.
These remarks will prove to be basic in the sequel.

It is clear that in any further consideration regarding metric
motions,
one has to add new objects into the system besides the metric tensor itself.
One possibility in order to define such
general concept
is to write down the expression of a particular action of a metric motion,
in the general form,
\beq
\label{eq-qdef}
\varphi^* g \equiv Q ,
\eeq
and impose now some restrictions on $ Q $
(if no restrictions on $ Q $ where
considered, Eq.~\rf{eq-qdef} would be nothing but an identity).
For instance, one has
$ Q = g $ in the case of isometries,
$ Q = e^{2 \Phi}\,  g $ in the conformal
case, $ Q = g + 2H \, \ll $, for \ks motions, and similarly for
the {\sl tt}-like motions ---those should be viewed as guiding examples.
The way one can restrict $ Q $ is varied (see \cite{ibragimov,muppi,bel98}).
It depends on the problem one wants to deal with.
We have taken into account several situations, and peculiarities, coming
from well known metric motions and collineations and their restrictions.
Here we will focus
only on the type of restrictions that can be expressed in terms
of geometrical objects, such as $ \lh $, $ \uh $, $ g $, etc.

\subsection{Definition of a generalized metric motion}
\label{ss-dgmg}

Bearing in mind the preliminary considerations of the previous section, we
shall proceed with a suitable definition of these generalized metric motions.
First, we must distinguish between
different classes of metric motions.
\bdf[Equivalence of metric motions]
\label{def-equiv}
Two metric motions, $ Q_1 $ and $ Q_2 $, will be
regarded as of the same class, or
equivalent, whenever their components (in any given basis) satisfy the smae restrictions, and this will
be denoted by $ Q_1 \simeq Q_2 $.
\edf

It is worth remarking that a precise definition
for each possible case is
meaningless, since one should list a lot of
different new situations which
may be described within a general framework.
Our aim is half-way
between the free-basis language of $ Q $, that does not give any specific
equations, and the particular examples of e.g., \ks motions, which are too
specific since only one geometrical object is used. In fact, there is
{\em no single} solution to this problem. It is rather a matter of
convention  that depends on the degree of generality one is ready to allow. Thus, the following approach can only be evaluated from the preceding physical aims and the results of the forthcoming sections.

On the other hand, since the language of metric motions can be described in
tensorial
terms,
one can consider a given cobasis, say $ \{ \ofo \} $
($ \Omega = 0, \ldots, n-1 $) in order to describe a {\em generalized metric
motion}.
It could be a natural tetrad, $ dx^{\alpha} $, but it is not always
the most suitable choice, as it can be already noticed from
Sect.~\ref{s-ksm} or from I, and Sect.~\ref{ss-lglm} later. \footnote{The
option of a natural cobasis will prove
to be useful mainly when trying to integrate the differential equations
of a particular metric group, see \cite{ibragimov,bel98}
and also Ex.~\ref{exlm} later.}
The cobasis formalism simply
introduces an arbitrary cobasis in the   representation of $ Q $. This, by
itself, does not introduce anything new  with respect to a general approach.
But it will allow us to
get the set of conditions of the Kerr-Schild, or
{\sl tt}-like examples,  for a general combination of geometric
objects in $ Q $.

In the case of isometric and conformal motions, this procedure is
clearly unnecessary. However, for
the rest, it is necessary. In most situations, one will have the bonus that
the very fundamental objects will constitute part of a tetrad.
For instance, \ks motions are a very good example of this simplification,
or, also,
in the {\sl tt} case, $ \uh $ may be chosen as
an element of the tetrad. Consequently, in most
situations, Sect.~\ref{s-mmgd},
the general framework will become quite natural.

We have also
studied other possibilities, but they have proven to be
equivalent to the formalism based on cobasis which, eventually,
appears to us as the most direct one. Other
approaches may be taken in problems with a different orientation
(see e.g., \cite{ibragimov}). We now introduce this formalism.

Let $ \cb $, $ \Omega = 0 $, $ \allowbreak\ldots $,  $ n-1 $,
be any given cobasis of a $ n $-dimensional
Riemannian manifold, and let $ \varphi_t $
represent a given set of diffeomorphisms,
with $ t \in (-a,a) $ with $ a \in {\bf R} $, $ \varphi_{t=0} = Id $.
Their action on
the metric tensor, $ g $, can always be written as
\beq
\label{eq5}
\rea \bg \equiv {Q}_t    =  \fa \tol,
\eeq
where $ \fa $ will be some $ C^{\infty} $ functions. However,
in order to be able to compute the transformation of a transformation, one
needs also to know the action of the given set of diffeomorphisms on the
basis. This will be
\beq
\label{eq6}
\rea \ofo  =  (M_{t})_{\; \Lambda}^{\Omega} \ofl,
\eeq
  whith $ (M_{t})_{\; \Lambda}^{\Omega} $
some $ C^{\infty} $ functions. 
In fact (and this is our point) it is only necessary to give some
{\em generic} restrictions on $ Q_t $. For instance, we have shown in the
previous examples, including the two well-known of metric motions,
that restricting $ Q_t $ to be proportional to e.g., $ g $, or
to $ \ll $, \ldots $\, $ suffices to give a meaningful set of motions.
These impositions
cannot be made precise now because we want to leave an open
window for other possible combinations. Nevertheless,
one can actually impose that a specific type of metric motion
is being considered with the aid
of Def.~\ref{def-equiv}. Then, we can give an
expression for a set of diffeomorphisms to become a generalized metric
motion, as follows.
\bdf[Generalized Metric Motion, I]
\label{def-ms}
A 1-parameter set of local transformations 
$ \{ \varphi_t \} $ of $ V_n $, $ t \in (-a,a) $ for some
appropriate $ a \in {\bf R} $, is called a {\em generalized metric motion} if and only if
Eqs.~\rf{eq5},~\rf{eq6} satisfy
  \beq
\label{eq-def1}
(Q_{t+t'})_{\Omega \, \Lambda} :=
\bigl(Q_{t} [\varphi_{t'}(x)]\bigr)_{\Pi \, \Sigma}\,
(M_{t'})^{\Pi}_{\; \Omega} \,(M_{t'})^{ \Sigma}_{\; \Lambda}
\simeq (Q_{t})_{\ol}, 
\eeq
where $ Q_t $ is some given class of tensors for any $ t $, $ t' $, $ t+t' \in (-a,a) $.
\edf
Notice that in~\rf{eq-def1} Def.~\ref{def-equiv} is used.

Moreover, the functions $ (M_t)^{\Omega}_{\; \Lambda} $
are linked with $ \fa $ through
$$ \begin{array}{rcl}
\rea g & = & \rea \bigl( g_{\ol} \tol \bigr) = \bigl( g[\varphi_t (x) ])_{\Sigma \Pi}
(M)^{\Sigma}_{\; \Omega} (M)^{\Pi}_{\; \Lambda}  \tol \cr
& = & \fa \tol . 
\end{array} $$
This gives  an equation in order to obtain, partly, the expression of
$ (M_t)^{\Omega}_{\; \Lambda} $ for a given generalized
metric motion. Indeed, the knowledge of
$ (M_t)^{\Omega}_{\; \Lambda} $, for all $ \Omega $, $ \Lambda $, for a given class of metric motionsis
sufficient to fix
the problem of a given generalized metric motion. In Sect.~\ref{s-mmgd} a
detailed implementation of this definition and remark
will be carried  out.

Since the transformations are continuous, it is possible
to give the infinitesimal characterization of a
generalized metric motion. This is the issue in the next section.
\subsection{Differential version of a metric motion}
\label{ss-ivmg}

The problem of  a metric motion is generally much easier handled
if one takes focuses on the differential equations which define the
action of the whole motion.

A standard calculation, similar to the one performed
in I for \ks motions, shows that the differential version of expressions \rf{eq5}, \rf{eq6} is \footnote{The
---rank-two symmetric--- tensor $ q $ has not to be confused with the 1-form
$ \qh $.}
\beq
\label{eq-56}
\lie \bg \equiv {\qx} = \qxb \tol, \quad \lie \ofo =
\mh_{\xi}^{\Omega} = (m_{\xi})_{\; \Lambda}^{ \Omega} \ofl,
\eeq
where $ \xiv $ is the differential generator associated with
the differential action of the motion, and
\beq
\qxb \equiv {d \fa \over dt} \biggm|_{t=0}, \qquad
(m_{\xi})_{\; \Lambda}^{\Omega}
\equiv {d(M_{t})_{\; \Lambda}^{\Omega}\over dt}
\biggm|_{t=0}. \nnb
\eeq

Obviously, in order to have a vector space, the set
$ \{ \qxb \} $ must be homogeneous,
to include $ \xiv = \vec 0 $, and,
moreover, their possible functional relationship among its elements
must be, at most, linear.
Moreover, since $ Q_t $ ---$t\in(-a,a)$--- must have some restrictions, as remarked in the
previous section, so must $ \qx $ too. That is, Eqs. \rf{eq-56} should
be understood as meaningful for some restricted class of tensors $ \qx $.
One thus has (Def.~\ref{def-equiv} for
equivalent classes of motions is used)
\bdf[Vector fields of a generalized metric motion]
\label{def-ms2}
Let $ \cb $, $ \Omega = 0 $, \allowbreak\ldots, $ n-1 $,
be any given cobasis of $ V_n$.
The solutions to the system of equations
\beq
  \label{eq8}
\lie \bg = \qx = \qxb \tol,  \\
\label{eq9}
\lie \ofo = (m_{\xi})^{\Omega}_{\ \Lambda} \ofl,
\eeq
  will
be considered to be {\em the vector fields
of a given generalized metric motion} if and only if
they satisfy
\beq
\label{eq10}
q_{\xi\xi} \equiv \lie \qx = [\xi^{\lambda}
\partial_{\lambda} \qxb +  2 (\qx)_{\Omega \Sigma}
(m_{\xi})^{\Sigma}_{\; \Lambda}] \tol \simeq \qx.
\eeq
where $ \qx $ is a given class of rank-two symmetric
tensor fields with $ \qxb $ homogeneous
functions, which are related with each other at most
by linear relations.
\edf
In Eq.~\rf{eq10} we have made use of the fact that
$ \qxb = (\qx)_{\Lambda \Omega} $.

The previous definition could be summarized by saying that
$ \xiv $ is a vector field of a metric motion if
and only if $ \lie \bg $ and $ (\lie)^2 \bg $
are of the same type ---according to Def.~\ref{def-equiv}.
The remarks given after Def.~\ref{def-ms} are also valid now,
when taking the differential action of the metric group.
The inclusion of a cobasis is again secondary, but it gives within
our aim a closer definition to any
future practical problem,
which departs from the isometric or
conformal motions.

Let us see how the preceding definition works with some examples, although
it will become apparent after having considered more general cases in the
following sections.
For instance, in the case of isometric motions no need for a cobasis
description appears and $ \qx = q_{\xi \xi} =  0 $.
Therefore,
the system of equations in Def.~\ref{def-ms2} turns out to be
simply Killing's equations (see \cite{killing,schouten,yano}).

In the case
of conformal motions, one has $ \qx = 2 \phi_{\xi} g $ and $ q_{\xi \xi} =
2(\lie \phi_{\xi} + 2 \phi_{\xi}^2 ) g $ so that $ q_{\xi \xi} \simeq \qx $.

Another example is that of \ks motions. In
this case $ \qx = 2 h_{\xi} \ll $ and $ q_{\xi \xi}
= 2(\lie h_{\xi} + 2 m_{\xi} h_{\xi} ) \ll $ if and only if $
\lie \lh = m_{\xi} \lh $ which represent the necessary contribution of
Eqs.~\rf{eq9}.
Thus, $ q_{\xi \xi} \simeq \qx $ so that \ks motions
verify Def.~\ref{def-ms2}.

On the contrary, we will show that double \ks motions, defined
by $ \qx = 2 f_{\xi} \mm + 2 h_{\xi} \ll $, where $ \mh $ is
a null 1-form field satisfying $ \mh \cdot \lh  = -1 $, do
not give rise to a new metric motion in the sense
of Defs.~\ref{def-ms} or~\ref{def-ms2}. This also shows that
Defs.~\ref{def-ms} or~\ref{def-ms2} are indeed half-way
between a too general
treatment of metric groups and a too specific treatment valid
only for a few examples. The aforementioned impossibility of double
Kerr-Schild motions is proven in Sect.~\ref{s-mmgd}.

Now we will check whether, within our generality, each vector field
fulfilling Def.~\ref{def-ms2} gives rise to a local flow
on the metric that belongs to Def.~\ref{def-ms},
thereby completing the general framework.
This was clear in the examples of metric motion before.
\blm
\label{lm-stan}
$ (\lie)^{(n)} g \simeq \lie g $, for all $ n \ge 1$.
\elm
\pr From Def.~\ref{def-ms2}, the vector
field $ \xiv $ must satisfy:
\beq
(\lie)^2 g \simeq \lie g. \nnb
\eeq
  Whence, by induction, one has
\beq
(\lie)^{(n)} g \simeq (\lie)^{(n-1)} g \nnb
\eeq
for all $ n \ge 2 $. Or, equivalently,
\beq
(\lie)^{(n)} g \simeq \lie g , \nnb
\eeq
as it was claimed.

This can also be shown by noticing, at this stage,
one is {\em not} interested in a precise
expression of each action. For instance, in Def.~\ref{def-ms2},
we require $ \lie \qx $ and $ \qx $ to be equivalent, in the sense
of Def.~\ref{def-equiv}. This allows to extend the equivalence to any further order. \N
\bpr
\label{prop-xf}
Any vector field $ \xiv $ belonging to Def.~\ref{def-ms2}
gives rise to a local 1-parameter group of local metric transformations
belonging to Def.~\ref{def-ms}.
\epr
  \pr For any particular $ \xiv $ one can construct the following
transformation
\beq
\rea g \equiv g + (\lie g) t + \cdots
+ {1 \over n!} \Bigl[(\lie)^n g\Bigr] t^n + \cdots \nnb
\eeq
where $ t \in (-a,a) $ for some appropriate $ a \in {\bf R} $.
Because $ t $ is only a parameter, from the preceding lemma, one has
\beq
[ (\lie)^{(n)} g ] t^n \simeq [\lie g] t . \nnb
\eeq
Whence, for any $ t \in (-a,a) -\{ 0 \} $, one gets:
\beq
\varphi_t^* g - g \simeq \qx = \lie g, \nnb
\eeq
which tells that $ \rea g $ is of a definite type for all $ t \in (-a,a) $, $ t \ne 0 $.

Consequently, $ \varphi^*_{t''} g
\simeq \varphi^*_{t'} g \simeq \rea g $ for any triad $ t'' $, $ t' $, $ t
$, where $ \varphi^*_t $, $ \varphi^*_{t'} $ and $ \varphi^*_{t''} $ are well defined.
In particular, for $ t'' = t+t' $. Therefore, $ \rea g $ constructed from the
vector fields $ \xiv $ give rise to a local 1-parameter group of local metric
motions
belonging to Def.~\ref{def-ms}. \N

A typical goal when dealing with systems
of differential equations is to obtain a normal
system for the unknowns
and for each particular metric group. Sometimes (almost
always, as we will show) this
will be impossible. The system
of Eqs. \rf{eq8}--\rf{eq10} turns out to be open
in most cases and infinite dimensional algebras appear,
see Sects.~\ref{s-ksm},~\ref{s-mmgd}
later, and also I. Therefore,
one can only
write it as close as possible to the claimed goal. We give the expressions
in App.~\ref{app-b}.

\section{Metric motions defined by different substructures}
\label{s-mmgd}

Hitherto isometric, conformal and \ks motions
are the only metric motions considered. After the introduction
of \ks motions and its study, it would be logical
to try to extend the number of candidates for metric
motions. The first natural path to generalize the three previous cases
of metric groups, in the spirit of Sects.~\ref{ss-rcpa},~\ref{ss-pfgm}, is
to consider some combinations of the most common
geometrical objects, e.g., vector fields (or either 1-forms).
Therefore, our aim here is mainly to implement
Defs.~\ref{def-ms},~\ref{def-ms2} and show how they work.
Yet we will spend some time to include the most
remarkable features of the candidates,
specially in Sect.~\ref{ss-lglm}.

For the sake of brevity, and due to their possible
physical applications, the study is centered
on four-dimensional spacetimes,
though many of the examples are also valid
---or can easily be extended--- to other
dimensions and signatures of the manifold, provided
these are compatible with the existence of each
geometrical element used.

First, backed by the results of \ks motions, we shall
try to extend its study in order to include another
(real-valued) null 1-form,
say $ \mh $ into the scheme (in this case, obviously, the metric
tensor cannot have an Euclidean signature). The two 1-forms
must satisfy $ \lh \wedge \mh \neq 0 $,
so that each 1-form is independent
from the other. Without loss of generality both null 1-forms
can be taken to satisfy the orthogonality condition
$ \lh \cdot \mh = -1 $ (signature $ +2 $). In this sense one
would obtain the usual expressions for ``advanced'' and
``retarded'' null 1-forms in any spacetime. In case
they are geodesic, the algebras can also be understood as the
Lie algebras generated by the geometrical structure of the light cone, a
point worth to be noticed.
Moreover, two such null vectors can be taken as elements
of any cobasis, and their study covers,
in some sense, half of the cases. This is the set
developed in Sect.~\ref{ss-lglm}.

Second, the other half is no longer expressible
in terms of other real-valued null 1-forms. In order to
complete the basis of the spacetime, one should include either
complex valued null 1-forms, or choose two spacelike
1-forms, say $ \ph $, $ \qh $, satisfying $ \ph \wedge \qh \wedge \lh \wedge
\mh \neq 0 $.
Hence our efforts after Sects.~\ref{ss-lglm},~\ref{ss-lgpq},
are centred in the study of candidates for metric groups
generated by this latter pair
(this section could be adapted to Euclidean signature).
The case of
metric groups that are generated by the combination
of a timelike and a spacelike 1-form are indeed equivalent to
Sect.~\ref{ss-lglm}. However it is worth
giving the solution, at least, in App.~\ref{app-c}. The reason is to be
found in the literature. In Sect.~\ref{ss-rcpa} it turned out that in aiming
to find new {\it Ans\"atze} from
1-form fields, different
to the GKS case, two major approaches have been taken in the literature.
One of them is explicitly based in null 1-forms,
whilst the other allows to include spacelike and timelike.
Let us recall that in our case they
are {\em internal groups of transformations} and our results complement, in
fact, those approaches in
an analogous
way as \ks motions supplement a new feature to the GKS {\it Ansatz}. Further
work has to be done
in this direction before one could centre the study in one of the two.
Therefore
it may be useful to have both expressions of the algebras in a summary,
one for each line of work  (moreover, App.~\ref{app-c} will
serve as well as a test of consistency for the whole scheme).
All these combinations clearly represent the seeds for any
other more general option and appear to be the simplest
ones concerning their geometrical structure, as stressed
also in Sect.~\ref{ss-rpa}.
\subsection{Metric motions generated by $ \ell ${--}$ m $}
\label{ss-lglm}

Let $ \lh $ and $ \mh $ be two given real null 1-form fields satisfying
$$ \lh \cdot \lh = \mh \cdot \mh = 0, \quad \lh \cdot \mh = -1 . $$

The differential expression
of a metric motion generated by these
two ingredients would be\footnote{A study based upon the finite
action of each motion is actually possible. However
the calculations are longer and more cumbersome, and they
do not contribute, generally, to a better understanding
of the basic features of each case. In the sequel, as no confusion can
arise, we avoid writing the subindex $ \xi $ in order to clarify the
notation. Moreover, in all the cases considered in this work it is quite easy to show that the set of solutions to each type of generalized metric motion gives rise to a vector space, and furthermore to Lie algebras. the steps are very similar to those of \ks motions, see I. Therefore, we omit their proofs here.}
\beq
\label{qlm}
q = \lie \bg = 2h \ll + c(\slm) + 2f \mm.
\eeq
In this scheme $ \bg $, $ \lh $, and $ \mh $ may be regarded as
data. In fact,
since $ \lh $ and $ \mh $ are null, only their direction is a necessary
datum, see also Sect.~\ref{s-ksm}. On the other side,
$ h $, $ f $, $ c $
are unknown $ C^{\infty} $ functions of the manifold yet to
be determined by the system of equations themselves, if possible. And
$ \xiv $ are the differential generators of the metric motion, which
constitute the main unknowns of the problem.

Taking $ \{ \lh, \mh \} $ as a part of the cobasis of spacetime,
the previous expression tells us that the relations among
the $ q_{\ol} $ are
rather simple: free weights in the portion expanded by $ \lh $
and $ \mh $, and zero-valued weights for the rest of the elements
that complete the cobasis.

Following Sect.~\ref{s-gmm}, the next
step is to calculate the expression of 
``$ \lie (\hbox{co-}\allowbreak\hbox{basis}) $''.
Whence, one will get the internal freedoms coming from the
local Lorentz character of the manifold. After that step, one
should check the stability property, in order to assure that
one deals with a coherent candidate for metric motion. The
cobasis will be completed with the addition of two
spacelike 1-forms, $ \ph $ and $ \qh $, satisfying
$ \ph \cdot \ph = \qh \cdot \qh =1$, $ \ph \cdot \lh = \ph \cdot \mh
= \qh \cdot \lh = \qh \cdot \mh = \ph \cdot \qh = 0 $, but
being otherwise arbitrary.

Since we are concerned with metric motions
{\em exclusively} generated by the pair $ \lh ${-}$ \mh $,
the rest of the cobasis only adds an isometric action to the
scheme. Therefore, the general scheme could be partially
restricted to the action upon the pair $ ( \lh, \mh ) $ if
one would like to focus on particular solutions.
Here, we shall develop the full method in order to clarify
the previous sections as much as possible.
These will be the actual steps and also for the following sections.
We summarize the results of this section in (A' stands for $ \lie A$)
\bpr[Metric motions generated by two null 1-forms]
\label{prop-lm}
The conditions in \allowbreak order
to have a generalized metric motion
generated by two given null 1-form fields, $ \lh $
and $ \mh $, i.e., Eqs.~\rf{qlm}, are
\beq
\label{eq13}
\cases{\lh' = \alpha_0 \lh - f \mh + \alpha_1 \ph + \alpha_2 \qh,
\quad \mh' = -h \lh - (c + \alpha_0) \mh + \beta_1 \ph + \beta_2 \qh, \cr
\ph' = \beta_1 \lh + \alpha_1 \mh + \gamma_1 \qh,
\quad \qh' = \beta_2 \lh + \alpha_2 \mh - \gamma_1 \ph,}
\eeq
with
\beq
\label{eq14}
\cases{ 2 h \alpha_1 + c \beta_1 = 0, \quad 2 f \beta_1 + c \alpha_1 = 0,
\cr
2 h \alpha_2 + c \beta_2 = 0, \quad 2 f \beta_2 + c \alpha_2 = 0,}
\eeq
and
\beq
\label{eq15}
\begin{array}{c}
\lie^2 g = 2 {\tilde h} \ll + {\tilde c} (\slm) + 2 {\tilde f} \mm,  \\
 \cases{
{\tilde h} = h' + 2 h \alpha_0- h c, \cr
{\tilde f} = f' - 2 f \alpha_0 - 3 f c, \cr
{\tilde c} = c' - 4 h f - c^2.}
\end{array}
\eeq
The volume element, $ \eta \equiv - \lh \wedge \mh \wedge \ph \wedge \qh $,
transforms according to
\beq
\label{vol-lm}
\eta' = - c \, \eta .
\eeq
\epr
In these expressions, $ \{ \lh $, $ \mh $, $ \qh $, $\ph \} $ is
{\em any} semi-null cobasis of the manifold and
$ \{\alpha_0 $, $ \alpha_1 $, $ \alpha_2 $, $ \beta_1 $,
$ \beta_2 $, $\gamma_1\} $
are $ C^{\infty}$ functions (we
will study their freedom below).

\pr Under any differential action of any metric
motion, one can write
\beq
\cases{
\lie \lh  =   A_1 \lh + A_2 \mh + A_3 \ph + A_4 \qh , \
\lie \mh  = B_1 \lh + B_2 \mh + B_3 \ph + B_4 \qh , \cr
\lie \ph  =  C_1 \lh + C_2 \mh + C_3 \ph + C_4 \qh, \
\lie \qh  =  D_1 \lh + D_2 \mh + D_3 \ph + D_4 \qh,} \nnb
\eeq
where the set of functions $ A_{\alpha} $, $ B_{\alpha} $, $ C_{\alpha} $, $
D_{\alpha} $ is,
for the moment, arbitrary. On the other hand, $ \bg $ may be
written as $ \bg = - (\slm) + \spq $,
because $ \{ \lh, \mh, \qh,\ph \} $ is a semi-null
cobasis. Then, calculating its Lie derivative and imposing
Eqs.~\rf{qlm}, one gets
\beq
\label{eq16}
\cases{
\lie \lh  =   A_1 \lh - f \mh + A_3 \ph + A_4 \qh , \cr
\lie \mh  = -h \lh - (c + A_1) \mh + B_3 \ph + B_4 \qh , \cr
\lie \ph  =  B_3 \lh + A_3 \mh + C_4 \qh, \cr
\lie \qh  =  B_4 \lh + A_4 \mh  - C_4 \ph.}
\eeq
This result assures the fulfillment of Eqs.~\rf{eq8}--\rf{eq9}.
The remaining ones, Eqs.~\rf{eq10},
contain the last conditions to be imposed. They assure
that $ \bg'' \simeq g' (=q) $ for $ {q} = 2 h \ll + c(\slm) + 2f \mm $.
In operative terms, it amounts to imposing
$ \bg'' = 2 {\tilde h} \ll + {\tilde c}(\slm) + 2{\tilde f} \mm $,
where the weights with a tilde represent a new set.
With the aid of Expr.~\rf{eq16}, one can then calculate
$ \bg'' $ and after imposing Eqs.~\rf{eq9}, one readily gets
conditions~\rf{eq13}--\rf{eq15} (where we have reordered
the names of the remaining functions).

Finally, it is now straightforward to calculate
$ \lie \eta $, being $ \eta $ the volume element of the
manifold. The result is Eq.~\rf{vol-lm}. \N

The study of Eqs.~\rf{qlm} should not
be restricted to its general expression. Equally
important are the subcases associated with further restrictions
on the weights $ h $, $ f $, $ c $ (see table~\ref{tb-lm}
and Prop.~\ref{prop-dis-lm} below). Therefore, we
shall
develop a study for all the possibilities. In particular, the
homogeneous case $ h= f = c = 0 $ lets $ \alpha_0 $,
$ \alpha_1 $, $ \alpha_2 $, $ \beta_1 $, $ \beta_2 $,
$ \gamma_1 $ free, which correspond to the generators of the
the local Lorentz algebra, as is expected.

Whereas the isometric motions can be completely recovered from the conformal
ones,
without losing generality, so that one can say that isometries are a subcase of conformal motions, it is no more valid in other general
situations. Actually, one has, from Prop.~\ref{prop-lm}
(see also the following sections)
\bpr
\label{prop-dis-lm}Isometries and the cases I and $ V_2$
are not subcases of the general solution, case $ V_1$.
\epr \N

This proposition stresses the fact that the vector fields of these 
three subcases cannot be recovered from the solution of 
the most general case by
imposing on the general case, $ V $, their particular
conditions, e.g., $ \!I_{(h \ne 0)} \neq {V_1}_{(f=c=0)} $,
or $ II_{f=0} \neq I_a $. Moreover,
these examples clearly show the role played by the form
---class--- of $ q $.
Fig.~\ref{fig-lm} shows the interrelation among the different cases.
Prop.~\ref{prop-dis-lm} is proven in the following.
\begin{table}
$$ \begin{array}{||c||l||c||}
\hline
{q} & \! {\rm Case} & \!\!\!\hbox{Extra freedoms} \cr
\hline
\hline
2h\, \ll & \! I_a & 4 \cr
\hline
2f \, \mm &\! I_b & 4 \cr
\hline
2h \, \ll + 2f \, \mm & II & \not\!\exists \cr
\hline
c \, (\slm) &\! III & 2 \cr
\hline
2 h\, \ll + 2c \, (\slm) & \! IV_a & 2 \cr
\hline
2 f \, \mm + 2c \, (\slm) &\! IV_b & 2 \cr
\hline
\begin{array}{l} \! 2h \, \ll + 2f \, \mm \cr + c \,  (\slm) \end{array} &
\begin{array}{l}\! \!\!\!V_1: c^2 \ne 4 h f
\cr \!\!\!\! V_2: h \!=\! a^2 f, c \! = \! \pm 2a f \end{array} &
\begin{array}{l}\hspace{-.5cm} V_1: 2 \cr \hspace{-.5cm} V_2: 3 \end{array}
\cr
\hline
\end{array} $$
\caption{\label{tb-lm}
Generalized metric motions generated by $ ( \lh,\mh ) $.
Note the presence of extra degrees of freedom in all the
cases that yield a metric motion. The non-existence of double
\ks motions is a remarkable result.  $ a $ is a constant under
the action of the motion $ V_2 $. (See the text for a detailed discussion.)}
\end{table}
\begin{figure}
$$ \begin{array}{|c|l|}
\hline
V_1 &\rightarrow IV \rightarrow III \cr
\hline
V_2 &\rightarrow {(V_2)}_{\, V_1}\cr
\hline
I &
\begin{array}{l}
\nearrow I_{V_1}, I_{IV} \cr
\searrow I_{V_2} \end{array}\cr
\hline
{\rm Isom} &\rightarrow Isom_{\rm I}
\begin{array}{l}
\nearrow {\rm Isom} _{\,V_1, IV, III}\cr
\searrow {\rm Isom}_{\, V_2} \end{array}\cr
\hline
\end{array} $$
\caption{\label{fig-lm}Interconnection among Lie algebras generated
by two null 1-form fields. Each level represents an independent algebra.
Each
arrow leads to a subalgebra. $ A_{B\ldots} $ means the restriction
of cases $ B $, \ldots to the conditions of case $ A $.
The last level are isometries. (See the text and table~\ref{tb-lm} for
details.)}
\end{figure}

\proclaim \ks motions.

\ni \ks motions, cases $ I_a $ and $ I_b $ in table~\ref{tb-lm},
correspond to setting $ c $ equal to zero and either $ h $ or $ f $
to be zero (clearly both situations are equivalent). The result is ($ h \ne
0 $)
$$\displaylines{
\cases{\lh' = \alpha_0 \lh, \quad
\mh' = - h \lh - \alpha_0 \mh + \beta_1 \ph + \beta_2 \qh, \cr
\ph' = \beta_1 \lh + \gamma_1 \qh, \quad \qh' = \beta_2 \lh - \gamma_1 \ph, }
\qquad{\tilde h} = h' +2h \alpha_0, \qquad \eta' = 0.} $$
There are four freedoms, namely $ \alpha_0 $, $ \beta_1 $,
$ \beta_2 $, and $ \gamma_1 $, which represent a freedom
in the parametrization of the null curves, defined by $ \lv $,
two boosts, in the plane $ \lh $--$ \ph $ and $ \lh $--$ \qh $,
and a rotation in the ``orthogonal'' plane $ \ph $--$ \qh $. The
freedom in $ h $ is not constant, but is actually a {\em functional}
freedom, see e.g., Sect.\ref{s-gld0} or~Ex.~\ref{exlm} later on. This was a
{\em new}
result for metric symmetries. It implies, for instance, that $ h $ cannot
always be
isolated in terms of the data, the metric tensor, the direction of $ \lh $,
and $
\xi_{\alpha} $, $ \xi_{\alpha \beta} $, \ldots and
tells us that the system is {\em open} (recall Prop.~\ref{open}).

\proclaim Double \ks motions.

\ni This case provides a very good example of how
some forms of $ {q} $
do not give rise to a new metric motion.

In this case one has to impose $ c= 0 $ and $ f  h  \neq 0 $.
However it turns out that condition~\rf{eq15} imposes
$ h  f  = 0 $ and, therefore, one gets a contradiction.
But $ c = h f = 0 $ leads
us to ordinary \ks groups! We thus get 
the announced result
\bpr
\label{prop-2ks}
{\it No} proper double \ks motions exist.
\N
\epr

This is the first example of a $ {Q} $ family that does not yield a
new metric group. Moreover, it shows that even though it
comes from the known double \ks
relation, a direct generalization of its predecessor, i.e.,
\ks relations, it completely departs from it as internal
transformations.
Another consequence is that any double \ks relation gives rise
to new ---non-equivalent--- metric tensors, in the sense described in
Sect.~\ref{ss-rcpa}.

Again, the study above
can be carried out working entirely with the finite action of
the motion. This study is generally more tedious. The procedure
is rather similar as before. Starting from the decomposition of 
$ g $ in the
semi-null cobasis, one finds the general expressions of
$ \rea (cobasis) $ that fulfill the equation
$ \rea (g) = g + 2 H_t \ll + 2 F_t \mm $. Then one imposes
Eqs.~\rf{compo} taking into account the expressions
obtained
for $ \rea (\lh ) $ and $ \rea ( \mh ) $. Whence, one obtains a condition
in order to eliminate the
crossed term $ \slm $. Eventually, one arrives at the same
conclusion as before. Each of the finite steps can be translated
into the infinitesimal version to see how each method works.
However, for the sake of brevity, we will not write down here
all these expressions.

\proclaim Semi-conformal motions.

\ni This new case is obtained setting
$ h = f = 0 $. The name semi-conformal is used
to mean that half of the cobasis is not used, so
that one cannot talk of $ c (\slm )$ as the usual conformal
factor. Instead it should be considered as a conformal problem
restricted to the subspace expanded by the pair
$ \lh ${-}$ \mh $ only. This point of view also points out
to how generalized metric motions allow new interesting
restrictions of the typical
conformal symmetry, {\em without}
reducing the dimension of the manifold, yet maintaining at the same time
the whole set of elements of a manifold.
This feature will be considered elsewhere. Ex.~\ref{exlm} will show another
interesting aspect connected with this motion.

In this case the conditions reduce to
$$\displaylines{
\cases{\lh' = \alpha_0 \lh, \quad \mh' = -(c+\alpha_0) \mh, \cr
\ph' = \gamma_1 \qh, \quad \qh' = - \gamma_1 \ph ,}
\quad {\tilde c} = c' - c^2, \quad \eta' = -c \, \eta.
} $$
Only $ \alpha_0 $ and $ \gamma_1 $ are freedoms
of the system. Their interpretation has been given elsewhere.
Again there are functional freedoms, as will be seen in
Ex.~\ref{exlm}.

\proclaim Semi-conformal \ks motions.

\ni Cases $ IV_a $ and $ IV_b $ are clearly equivalent. The remarks
of the former section are again valid here. This case corresponds
to setting either $ h=0 $ or $ f=0 $. For the case $ f = 0 $,
$ h  c \ne 0 $, Eqs.~\rf{eq13}--\rf{vol-lm}
translate into
$$ \displaylines{
\cases{
\lh' = \alpha_0 \lh, \quad \mh'= - h \lh - (c+\alpha_0) \mh, \cr
\ph' = \gamma_1 \qh, \quad \qh' = \gamma_1 \ph,}
\quad  \cases{\tilde h = h' + 2 h \, \alpha_0 - h \, c, \cr
\tilde c = c' - c^2,} \quad \eta' = - c \, \eta.}
$$
The other situation, i.e., $ h = 0 $, $ f c \ne 0 $ is analogous,
changing $ h $ by $ f $.

\proclaim Semi-conformal double \ks motions.

\ni This last situation is the most general one within our scheme.
Their conditions are the ones displayed in Prop.~\ref{prop-lm}.
Before proceeding with their analysis, we would like to introduce
these motions from a more intuitive point of view that trivializes
some of their consequences.

Since  $ \bg = -(\slm) + \bg_{\perp} $, where $ \bg_{\perp} $
is the reduction of the metric tensor to the rest of the cobasis,
Eqs.~\rf{qlm} can be written as $ \lie (\slm ) \! = \lie \bg_{\perp} + 2
{\tilde h} \ll + 2 {\tilde c}(\slm) + 2 {\tilde f} \mm $,
where $ {\tilde f} $, $ {\tilde c} $, $ {\tilde h} $ are respectively
$ -f $, $ -c $, $ -h $. Let us further assume that $ \lie \bg_{\perp} $
is zero, i.e., that $ \xiv $ are isometries of the complementary subspace.
This last condition is not trivial, but it is not difficult to find many
spacetimes fulfilling it (e.g., decomposable spacetimes).
Then, the restricted problem
reduces almost to an identity. Therefore, the equations are
almost equivalent to a trivial
problem on metric groups in the Riemannian submanifold
generated by the pair $ \lh ${-}$ \mh $, and a great variety
of solutions are expected.

Despite these conclusions, let us carefully analyze conditions~\rf{eq14}.
One must distinguish between two possibilities: $ c^2 \neq 4 h  f $, case
$ V_1 $, and $ c^2 = 4 h f $, case $ V_2 $.

In case $ V_1 $, one obtains $ \alpha_1 = \alpha_2= \beta_1 = \beta_2 = 0 $.
The
complete solution is
\beq
\cases{\lh' = \alpha_0 \, \lh - f \, \mh, \quad
\mh' = - h \, \lh - (c + \alpha_0) \, \mh, \cr
\ph' = \gamma_1 \, \qh, \quad \qh' = - \gamma_1 \, \ph,} \nnb
\eeq
and, where $ {\tilde h} $, $ {\tilde f} $, $ {\tilde c} $ and $ \eta' $ are
given in Eqs.~\rf{eq16}--\rf{vol-lm}.

The second case, $ V_2 $, is interesting because, apparently, it introduces
a non-linear
relation among the components of $ q $.
However, one
knows that they must be linear in order to form a vector space.
For instance, imposing $ (c_{A+B})^2 = 4 (h_{A+B}) \, (f_{A+B}) $,
where $ c_{A+B} = c_A + c_B $, $ h_{A+B} = h_A + h_B $,
$ f_{A+B} = f_A + f_B $, i.e., the
weights of $ \xiv_{A+B} \equiv \xiva + \xivb $
---recall that, by hypothesis, $ c_A^2 = 4 h_A f_A $,
$ c_B^2 = 4 h_B f_B $---, one gets the result
$ h_A/f_A = h_B / f_B $ for any $ A $, $ B $. Therefore,
$ h / f = const. $ If $ h/f = const. $, and $ c^2 = 4hf $,
one gets either $ h= a^2 f $, $ c= \pm 2 a f $, where $ a $ is a constant,
or $ f = b^2 h $, $ c = \pm 2 b h $, where $ b $ is a constant. Both options
are
equivalent except when $ a =0 $ or $ b= 0 $. Despite this possibility, the
results
of any case are analogous to each other. Therefore, we will consider in the
sequel
the first situation, i.e., $ f $ free. Notice
that ``const.'' stands here for {\it constant under the
action of the motion}. Thus, it can be a function built
upon invariants of the motion. Continuing, $ {q} $ may
be reordered to be expressed as
$$ {q} = (\mp 4 a f) \rh_{\pm} \otimes \rh_{\pm}, $$
where $ \rh_{\pm} \equiv (\sqrt{a/2})(\lh \, \pm \, \mh/a) $, with $
\rh_{\pm} \cdot \rh_{\pm} = \mp 1 $ and $ \rh_{\pm} \cdot \rh_{\mp} = 0 $. $
\alpha_1 $,
$ \beta_1 $ are now free. This algebra formed by two spacelike, timelike,
1-forms will be delayed until we present the Lie
algebras generated by $ \ph $--$\qh $ and $ \uh $--$ \nh $,
respectively.

We shall now give the first explicit solutions to the previous
algebras. We shall deal with flat spacetime ($ n =4 $)
and $ \lh $, $ \mh $
covariantly constant.\footnote{Although the generalization to an
$ n $-dimensional flat metric is straightforward,
it is not essential in the present discourse.}
Despite the apparent simplicity, it will display {\it all} the
properties of the algebras, as, e.g., the structure given
in Fig.~\ref{fig-lm}. We recall that the two independent
covariantly constant null 1-forms of flat spacetime can be chosen,
without losing generality, as the ``advanced'' and ``retarded''
parallel null 1-forms. Thus, the algebras can be considered as
generated by the geometrical structure of the light cone of flat
spacetime.
In Cartesian coordinates, one has
$ ds^2 = -dt^2 + dz^2 + dx^2 + dy^2 $, and $ \lh $ and
$ \mh $ may be written as $ \lh = \dx u $, $ \mh = \dx v $,
where $ u \equiv (1/\sqrt{2})(t-z) $, $ v \equiv (1/\sqrt{2})(t+z) $.
\bex
\label{exlm}
In flat spacetime and for the advanced and retarded null 1-forms,
the vector fields of the metric motions generated by them are-
\begin{enumerate}
\item For the case $ V_1 $,
\beq
\label{eq-v1}
\xiv  =  A\partial_u + B \partial_v +
(c_0 \, y + d) \partial_x + (-c_0 \,x +e) \partial_y   ,
\eeq
where $ A $, $ B $ are {\it arbitrary} $ C^{\infty} $ functions of $ u $,
$ v $, and $ c_0 $, $ d $, $ e $ are constants. Moreover,
\beq
\label{hfgv1}
h= - \partial_u B, \ f= - \partial_v A, \ g= - (\partial_v B +
\partial_u A),  \ \alpha_0= \partial_u A, \ \gamma_1= c_0 .
\eeq
\item For cases $ IV_a $, $ IV_b $ and $ III $, the previous one
restricted to each situation.
\item For the case $ V_2 $,
\beq
\label{eq-v2}
\begin{array}{rcl}
\xiv  &= & A\partial_{{\bar s}_{\pm}} + (b_1 x + b_2 y ) \partial_{{\bar
s}_{\mp}}
+ k \partial_{v} + (c_0 \, y + b_1 s_{\mp} + d_1) \partial_x \\
& &+ (-c_0 \,x + b_2 s_{\mp} + d_2 ) \partial_y   ,
\end{array}
\eeq
where $ A $ is an {\it arbitrary} $ C^{\infty} $ function of
$ s_{\pm}\equiv v \, \pm \, a u $, with $ a \neq 0 $ a constant,
$ {\bar s}_{\pm} \equiv u \, \pm \, a v $, and
$ b_1 $, $ b_2 $, $ k $, $ c_0 $, $ d_1 $, $ d_2 $,
are constants. Moreover,
\beq
f = - {\dot A} , \quad \alpha_1 = b_1, \quad \alpha_2 = b_2,
\eeq
where $ {\dot {()} } $ means total derivation with respect to $ s_{\pm} $,
and,
recall, $ h =a^2 f $, $ c= \pm 2 a f $, $ \alpha_0 = \mp a f $.
\item For the case $ I_a $ ($ I_b $ is analogous),
\beq
\label{eq-exks}
\xiv = A \partial_u + (-{\dot A} v + B + x {\dot C} +
y {\dot D}) \partial_v + (d y + C) \partial_x + ( - d x + D ) \partial_y
\eeq
where $ A $, $ B $, $ C $, $ D $ are {\it arbitrary} $ C^{\infty} $
functions of  $ u $, $ {\dot {()}} $ means derivation with respect to $ u $,
and $ d $ is a constant. Moreover,
\beq
\label{eq-hfg}
h = {\ddot A} v - {\dot B} - x{\ddot  C} - y {\ddot D}, \quad \alpha_0 =
{\dot
A},
\quad \beta_1 = {\dot C}, \quad \beta_2 = {\dot D}, \gamma_1 = d.
\eeq
\item Notice that, in any case, setting $ h = f = c = 0 $, one obtains
particular
sets of {\it restricted} isometries.
\end{enumerate}
All the Lie algebras, except the latter, are {\em infinite} dimensional.
\eex
\pr The line element of a 4-dimensional flat spacetime
can be written as
\beq
\label{a-ds2}
ds^2 = - 2 \, du \, dv + \, dx^2 + \, dy^2 .
\eeq
The two parallel null directions are, without lost of generality,
$ \lh = \dx u $ and $ \mh = \dx v $. First we must solve
\beq
\label{a-lm}
\lie \eta = 2 h \ll + 2 f \mm + c (\slm) ,
\eeq
where $ \eta $ is the flat metric. After this, we will impose the conditions
coming from Eqs.~\rf{eq9}, i.e., $ (\lie )^2 g \simeq \lie g $,
or {\it stability} conditions.

Consider a generic vector field $ \xiv $,
\beq
\label{a-xiv}
\xiv = \xi^{\lambda} \partial_{\lambda},
\eeq
where $ \lambda = u $, $ v $, $ x $ and $ y $.
Eqs.~\rf{a-lm} may be written as
\beq
\xi^{\lambda} \partial_{\lambda} \eta_{\ab}
+ \eta_{\alpha \lambda} \partial_{\beta} \xi^{\lambda}
+ \eta_{\beta \lambda} \partial_{\alpha} \xi^{\lambda}
= 2 h \, \delta^{u}_{\alpha}\delta^{u}_{\beta}
+ 2 f  \, \delta^{v}_{\alpha}\delta^{v}_{\beta}
+ c \, ( \delta^{u}_{\alpha}\delta^{v}_{\beta} + \delta^{u}_{\beta}
\delta^{v}_{\alpha}). \nnb
\eeq
Whence, using~\rf{a-ds2} and~\rf{a-xiv}, one gets
\beq
h = - \partial_u \xi^v , \nnb \\
f = - \partial_v \xi^u , \nnb \\
c = -( \partial_u \xi^u + \partial_v \xi^v ) , \nnb \\
\label{a-ux}
\partial_u \xi^x = \partial_x \xi^v , \\
\label{a-uy}
\partial_u \xi^y = \partial_y \xi^v , \\
\label{a-vx}
\partial_v \xi^x = \partial_x \xi^u , \\
\label{a-vy}
\partial_v \xi^y = \partial_y \xi^u , \\
\label{a-xy}
\partial_x \xi^x = \partial_y \xi^y = \partial_x \xi^y + \partial_y \xi^x =
0 .
\eeq
From Eqs.~\rf{a-xy} one readily gets
\beq
\xi^x = c_0 y + F_1(u,v), \quad \xi^y = - c_0 x + F_2 (u,v), \nnb
\eeq
where $ c_0 $ is a constant and
$ F_1 $, $ F_2 $ are $ C^{\infty} $ functions of their arguments.
Substituting this result in Eqs.~\rf{a-ux}--\rf{a-vy},
and using Eqs.~\rf{a-xy}, one gets
\beq
\xi^u =  \bigl[ F_3(u,v) + x \partial_v F_1(u,v) + y \partial_v F_2(u,v)
\bigr], \nnb \\
\xi^v =  \bigl[ F_4(u,v) + x \partial_u F_1(u,v) + y \partial_u F_2(u,v)
\bigr], \nnb
\eeq
where $ F_3 $, $ F_4 $ are $ C^{\infty} $ functions of their arguments.
Collecting all results, we obtain the general solution for Eqs.~\rf{a-lm}
\beq
\xiv & = &\bigl( F_3 + x \,\partial_v F_1 + y \, \partial_v F_2 \bigr)
\,\partial_u + \bigl( F_4 + x\, \partial_u F_1 + y\, \partial_u F_2 \bigr)
\,\partial_v \nnb \\
\label{a-xi}
& & + \bigl(c_0 \,y + F_1 \bigr) \partial_x +
\bigl( - c_0\, x + F_2 \bigr) \partial_y, \\
\label{a-h}
h & = & - \bigl( \partial_u F_4 + x \,\partial_{uu} F_1 + y\, \partial_{uu}
F_2
\bigr), \\
\label{a-f}
f & = & - \bigl( \partial_v F_3 + x \,\partial_{vv} F_1 + y \,\partial_{vv}
F_2
\bigr), \\
\label{a-g}
c & = & - \bigl( \partial_u F_3 + \partial_v F_4
+ 2x\, \partial_{uv} F_1 + 2y\, \partial_{uv} F_2 \bigr) .
\eeq
The next
step consists in imposing $ (\lie)^2 g \simeq \lie g $.
This is collected in Conds.~\rf{eq13},~\rf{eq14}
of Prop.~\ref{prop-lm}.

In order to do this, we first
calculate $ \lie \lh $, $ \lie \mh $, $ \lie \ph $
and $ \lie \qh $ for the general solution. We will take, without
losing generality, $ \ph = \dx x $, $ \qh = \dx y $.
The result is,
\beq
\cases{\lie \lh = \alpha_0 \lh - f \mh + \alpha_1 \ph + \alpha_2 \qh, \cr
\lie \mh = -h \lh - (c + \alpha_0) \mh + \beta_1 \ph + \beta_2 \qh,
\cr
  \lie \ph = \beta_1 \lh + \alpha_1 \mh + c_0 \, \qh,
\quad \lie \qh = \beta_2 \lh + \alpha_2 \mh - c_0 \, \ph,} \nnb
\eeq
where
\beq
\cases{\alpha_0 \equiv \partial_u F_3 + x \,
\partial_{uv} F_1 + y \, \partial_{uv} F_2,
\quad \alpha_1 = \partial_v F_1, \cr
\alpha_2 = \partial_v F_2, \quad
\beta_1 = \partial_u F_1, \quad \beta_2 = \partial_u F_2, } \nnb
\eeq
and $ f $, $ c $, $ h $ are given in Eqs.~\rf{a-h}--\rf{a-g}.
The imposition of Eqs.~\rf{eq14} clearly depends on each case.

1. For the case $ V_1 $, where $ c^2 \neq 4hf $,
$ \alpha_1 $, $ \alpha_2 $, $ \beta_1 $ and $ \beta_2 $ must vanish.
This yields $ F_1 = d = const. $ and $ F_2 = e = const. $ Consequently,
one obtains Eqs.~\rf{eq-v1}--\rf{hfgv1}.

2. For the cases $ IV_a $, $ IV_b $ and $ III $, Eqs.~\rf{eq14}
yield the same restrictions as before,
$ \alpha_1 = \alpha_2 = \beta_1 =  \beta_2 = 0 $.
Moreover, there is one further restriction
coming from $ f = 0 $, $ h = 0 $ and $ f = h = 0 $, respectively.
Therefore the solution is, in each case, a reduction of the
previous one.
This result also shows that the relation $ III  \subset IV_a \subset V_2 $,
or
$ III \subset IV_b \subset V_2 $,
follows the same pattern as well known metric
symmetries, i.e., ``Killing vector fields $ \subset $ homothetic vector
fields
$ \subset $  special conformal vector fields $ \subset $ conformal vector
fields''.

3. For the case $ V_2 $, one must impose
\beq
h = a^2 f , \quad c = \pm 2 a f , \nnb
\eeq
where $ a $ a constant, and
\beq
\alpha_0  = \mp a f, \quad
\beta_1 =  \mp a \, \alpha_1, \quad
\beta_2  =  \mp a \, \alpha_2 . \nnb
\eeq
The last two equations yield
\beq
F_1  = F_1 (v \mp a u), \quad F_2 = F_2 (v \mp a u) . \nnb
\eeq
Furthermore, since $ c + \alpha_0 = \pm a f $,
one has, using Eqs.~\rf{a-f}--\rf{a-g},
\beq
\label{a-gaf}
- \partial_v F_4 \, \pm\, a \partial_v F_3 \,\pm \, 2ax {\ddot F_1}
\, \pm \, 2 a {\ddot F_2} \, = \, 0,
\eeq
where we have put $ {\ddot F} \equiv d^2 F/ds_{\mp}^2 $ with
$ s_{\mp} \equiv v \mp a u $. As
$ F_3 $, $ F_4 $ are functions on $ u $ and $ v $, $ {\ddot F_2} $ and
$ {\ddot F_1} $ must vanish, that is,
\beq
F_1 = b_1 s_{\mp} + d_1, \quad F_2 = b_2 s_{\mp} + d_2 , \nnb
\eeq
with $ b_1 $, $ b_2 $, $ d_1 $ and $ d_2 $ constants.
Coming back to Eq.~\rf{a-gaf}, we get
\beq
\label{a-vf4}
  \partial_v F_4 = \pm a \partial_v F_3.
\eeq
  Similarly, from $ 2\alpha_0 +  c = 0 $, $ h = a^2 f $ and $ h = \mp a
\alpha_0
$,
one gets, respectively,
\beq
\label{a-3}
\partial_u F_3 = \partial_v F_4, \quad \partial_u F_4 = a^2 \, \partial_v
F_3,
\quad
\partial_u F_4 = \mp a \, \partial_u F_3 .
\eeq
Combining Eqs.~\rf{a-vf4}--\rf{a-3}, one easily gets
\beq
F_3 = F_3 (s_{\pm}), \quad F_4 = \pm a F_3 + k, \nnb
\eeq
where $ k $ is a constant.

Substituting the expressions for $ F_1 $, $ F_2 $, $ F_3 $ and $ F_4 $
into~\rf{a-xi} we get
\beq
\xiv = \bigl[ F_3(s_{\pm}) + b_1 x +  b_2 y \bigr]
\partial_u + \bigl[\pm a F_3(s_{\pm}) + k  \mp a (b_1 x + b_2 y) \bigr]
\partial_v \cr
+ \bigl[c_0 y + b_1 s_{\mp} + d_1 \bigr] \partial_x +
\bigl[ - c_0 x + b_2 s_{\mp} + d_2 \bigr] \partial_y. \nnb
\eeq
Reordering the components, and defining $ {\bar s}_{\pm}
\equiv u \pm a v $, one finally gets
\beq
\xiv = F_3(s_{\pm}) \partial_{{\bar s}_{\pm}} + (b_1 x  +b_2 y)
\partial_{{\bar s}_{\mp}} + k \partial_v + ( cy + b_1 s_{\mp} + d_1)
\partial_x \cr
+ ( - cy + b_2 s_{\mp} + d_2) \partial_y.
\eeq
This is the result given in Eqs.~\rf{eq-v2}, where the names
of the functions have been changed for convenience.

4. For the cases $ I_a $ or $ I_b $, the solution can be read from the
result given in I. The conditions
are (e.g., for the case $ I_a $): $ f= c = \alpha_1 = \alpha_2 = 0 $.

5. For the case where $ f = c = h = 0 $ is imposed from the
beginning, the vector fields are
the Killing vector fields of flat spacetime.
However, if $ f = 0 $, $ c = 0 $ or $ h = 0 $ are imposed
in the solutions of each of the preceding cases, one
obtains a subset of
Killing vector fields. This is worth to be remarked
since it shows how the Poincar\'e group
can be restricted by non-isometric groups. \N

It is not difficult from Prop.~\ref{prop-two} to enlarge Ex.~\ref{exlm} with several different solutions in other
spacetimes, e.g., for pp-waves or Vaydia spacetimes, and for other $ \lh $, $ \mh $. However, the previous example shows
very clearly all the basic considerations, and subtleties, of previous
sections.
For instance, the interconnection among the algebras
(Prop.~\ref{prop-dis-lm}) appears very clearly.
Another remarkable fact is that the Lie algebras are,
all but in the isometric case, infinite dimensional.
Therefore, their differential systems are open, and their integrability
equations are those of App.~\ref{app-b} (indeed, also used here in their simplest form). Nevertheless, there may be some, more or less, generic
conditions
under which each case may yield a closed system, and the algebra may
become finite dimensional in a natural way.

As of now, one can focus on their finite
dimensional subalgebras. This is
accomplished setting $ f $, $ h $ and $ c $ to be constants, say
$ f_1 $, $ h_1 $ and $ c_1 $.
For the case $ V_1 $, one gets
($ \lambda^1 = - f_1 $, $ \lambda^2 = - h_1 $,
$ \lambda^3 = - c_1 $)
\beq
\label{eqs-falm}
\begin{array}{lcl}
\xiv & = & \lambda^1 v \partial_u + \lambda^2 u \partial_v + \lambda^3 v
\partial_v
+ \lambda^4 (u \partial_u - v \partial_v) + \lambda^5 \partial_u + \lambda^6
\partial_v \cr
& & + \lambda^7 (y \partial_x - x\partial_y)
+ \lambda^8 \partial_x + \lambda^9 \partial_y
\equiv \lambda^{\Omega} \xiv_{\Omega},
\quad \Omega=1, \ldots, 9.
\end{array}
\eeq
The non-zero Lie brackets are
$$\begin{array}{lllll}
[\xiv_1 ,\xiv_2 ] = -\xiv_3 , &
\![\xiv_1 ,\xiv_3 ] = -\xiv_1 , &
\![\xiv_1 ,\xiv_4 ] = 2 \xiv_1 , &
\![\xiv_1 ,\xiv_6 ] = -\xiv_5 , &
\!\![\xiv_2 ,\xiv_3 ] = \xiv_2 , \cr
[\xiv_2 ,\xiv_4 ] = -2 \xiv_2 , &
\![\xiv_2 ,\xiv_5 ] = -\xiv_6 , &
\![\xiv_3 ,\xiv_6 ] = -\xiv_6 , &
\![\xiv_4 ,\xiv_5 ] = -\xiv_5 , &
\!\![\xiv_4 ,\xiv_6 ] = \xiv_6 , \cr
[\xiv_7 ,\xiv_8 ] = \xiv_9 , &
\![\xiv_7 ,\xiv_9 ] = -\xiv_8 , &
\end{array} $$
For the cases $ IV_a $, $ IV_b $ or $ III $, the corresponding restriction
applies. For
the \ks case, we refer the reader to I, where also different
finite dimensional algebras are given.
Finally, for the case $ V_2 $, one has ($ \lambda^1 = -f_1 $, $ a \neq 0 $)
\beq
\begin{array}{lcl}
\xiv & = & \lambda^1 s_{\pm} \partial_{{\bar s}_{\pm}}
+ \lambda^2 (x \partial_{{\bar s}_{\mp}} + s_{\mp} \partial_x)
+ \lambda^3 (y \partial_{{\bar s}_{\mp}} + s_{\mp} \partial_y)
+ \lambda^4 \partial_{{\bar s}_{\pm}}
+ \!\lambda^5 \partial_{{\bar s}_{\mp}} \cr
& & + \lambda^6  (y \partial_x - x\partial_y)
+ \lambda^7 \partial_x + \lambda^8 \partial_y
\equiv \lambda^{\Omega} \xiv_{\Omega},
\quad \Omega=1, \ldots, 8,
\end{array} \nnb
\eeq
and
$$\begin{array}{lllll}
[\xiv_1 ,\xiv_5] = \mp 2a \xiv_5, &
\![\xiv_2 ,\xiv_3 ] = \pm 2 a\xiv_6 , &
\![\xiv_2 ,\xiv_6 ] = - \xiv_3 , &
\![\xiv_3 ,\xiv_6 ] = -\xiv_2 , \cr
[\xiv_2 ,\xiv_7 ] = - \xiv_5 , &
[\xiv_3 ,\xiv_4 ] = - \xiv_5 , &
\![\xiv_6 ,\xiv_7 ] = \xiv_8 , &
\![\xiv_6 ,\xiv_8 ] = -\xiv_7 . &
\end{array} $$

Notice that by choosing $ c_1 $ equal to zero in case $ V_1 $,
one gets
$ \lie \eta = 2 h_1 \ll + 2 f_1 \mm $.
Therefore, it seems that one has found a counter-example
of our results about double Kerr-Schild. Nevertheless,
calculating $ (\lie)^2 \eta $, one gets
$ (\lie)^2 \eta = 4[ \alpha_0  h_1 \ll - h_1 f_1 (\slm) - \alpha_0 f_1 \mm ]
$.
Closer inspection
shows what happens: a semi-conformal term appears, so
that it is not a proper double \ks motion, but a special
case of semi-conformal double \ks groups. (Recall that in general
any metric group is defined within the two first
orders).

But there is something else interesting. We have been able to
start a problem without an {\it explicit} semi-conformal term
at first order. It has explicitly appeared only up to the next order.
This is a novelty with respect to all previously known metric
symmetries, including \ks symmetries; if a term does not
appear at the first order, it does not appear ever more. One
can now start conformal explicit actions at the second order
level, leaving the first order as another group action. A
point worth to have in consideration in case one wishes to
apply metric groups to perturbative works or some
processes in which a symmetry changes.

Finally, let us recall that the results obtained in this section
may be helpful in the study of finite metric relations based upon
two null vectors, \cite{ps} or with the conformal \ks
{\it Ansatz}, e.g., \cite{mh}.
\subsection{Metric motions generated by $ p ${--}$ q $}
\label{ss-lgpq}

In this section we will give the main results only.
Let $ \ph $ and $ \qh $ be two givwn 1-forms satisfying
$ \ph \cdot \ph = \qh \cdot \qh = 1 $, $\ph \cdot \qh = 0  $.
The differential expression
of a metric motion generated by these
two ingredients is
\beq
\label{qpq}
\lie \bg = 2h \pp + c(\spq) + 2f \qq. \nnb
\eeq
In this scheme $ \bg $, $ \ph $, and $ \qh $ may be regarded as data.
On the other hand $ h $, $ f $, $ c $
are unknown $ C^{\infty} $ functions of the manifold, and
$ \xiv $ are the infinitesimal generators of the group, which
constitute the major unknowns of the problem.

Taking $ \{ \ph, \qh \} $ as part of the cobasis of spacetime, it may be
completed with the addition of two
null 1-forms, $ \lh $ and $ \mh $, satisfying
$ \lh \cdot \lh = \mh \cdot \mh = 0 $, $ \lh \cdot \ph = \lh \cdot \qh
= \mh \cdot \ph = \mh \cdot \qh = \lh \cdot \mh + 1 = 0 $, but
being otherwise arbitrary (this also enables to compare this section with
the former in a direct way).
We summarize the results of this section as follows ($ A' $ stands for $
\lie A $)
\bpr[Metric motions generated by $ p $--$ q $]
\label{prop-pq}
The conditions in order
to have a generalized metric motion
generated by two given spacelike 1-form
fields, $ \ph $ and $ \qh $, i.e., Eqs.~\rf{qpq}, are
\beq
\label{eq18}
\cases{\lh' = \alpha_0 \lh + \alpha_1 \ph + \alpha_2 \qh,
\quad \mh' = - \alpha_0 \mh + \beta_1 \ph + \beta_2 \qh, \cr
\ph' = \beta_1 \lh + \alpha_1 \mh + h \ph + \gamma_1 \qh,
\quad \qh' = \beta_2 \lh + \alpha_2 \mh+ (c - \gamma_1) \ph + f \qh,}
\eeq
with
\beq
\label{eq19}
\cases{ 2 h \alpha_1 + c \alpha_2 = 0, \quad 2 f \alpha_2 + c \alpha_1 = 0,
\cr
2 h \beta_1 + c \beta_2 = 0, \quad 2 f \beta_2 + c \beta_1 = 0,}
\eeq
and
\beq
\label{eq20}
\cases{
{\tilde h} = h' + 2 h^2 + c ( c - \gamma_1) , \cr
{\tilde f} = f' + 2 f^2 +  c \gamma_1, \cr
{\tilde c} = c' + 2(h-f) \gamma_1 + (h + 3f) c.}
\eeq
The volume element, $ \eta \equiv - \lh \wedge \mh \wedge \ph \wedge \qh $,
transforms according to
\beq
\label{vol-pq}
\eta' = ( h + f ) \, \eta .
\eeq
\epr
In these expressions $ \{ \lh $, $ \mh $, $ \qh $, $\ph \} $ is
{\it any} semi-null cobasis of the manifold and
$ \{\alpha_0 $, $ \alpha_1 $,
$ \alpha_2 $, $ \beta_1 $, $ \beta_2 $, $ \gamma_1\} $
are $ C^{\infty}$ functions.

\pr The proof follows exactly the proof of Prop.~\ref{prop-lm}. \N

Again one has, from Prop.~\ref{prop-pq}
\bpr
\label{prop-dis-pq}Isometries and cases I, $ II_1 $, $ IV $ and $ V_2 $
are
not subcases of the general solution, case $ V_1 $. \N
\epr

\begin{table}
$$ \begin{array}{||c||l||c||}
\hline
{q} & {\rm Case} & \hbox{Extra freedoms} \cr
\hline
\hline
2h\, \pp & I_a & 3 \cr
\hline
2f \, \qq & I_b & 3 \cr
\hline
2h \, \pp + 2f \, \qq &
\begin{array}{l} II_1: h \neq f \cr II_2: h = f  \end{array} &
\begin{array}{l} II_1: 1 \cr II_2: 2 \end{array} \cr
\hline
c \, (\spq) & III & \not\!\exists \cr
\hline
2 h\, \pp + 2c \, (\spq) & IV_a & 1 \cr
\hline
2 f \, \qq + 2c \, (\spq) & IV_b & 1 \cr
\hline
\begin{array}{l} 2h \, \pp + 2f \, \qq \cr + c \,  (\spq) \end{array} &
\begin{array}{l} V_1: c^2 \ne 4 h f \cr V_2: h = a^2 f, c = \pm 2 a f
\end{array} &
\begin{array}{l} V_1: 2 \cr V_2: 3 \end{array} \cr
\hline
\end{array} $$
\caption{\label{tb-pq}Generalized motions generated by
$ (\ph, \qh ) $.
The non-existence of Lie algebras for the case $ III $
is a consequence of the stability property. $ a $ is
a constant under the action of the motion $ V_2 $. The latter
is equivalent  to the case I.
(See the text for a detailed discussion.)}
\end{table}
\begin{figure}
$$ \begin{array}{|c|l|}
\hline
V_1 &\rightarrow II_2 \cr
\hline
V_2  = I & \cr
\hline
IV_{V_1} & \rightarrow IV \cr
\hline
{II_1}_{\, V_1} & \rightarrow II_1 \cr
\hline
I & \rightarrow I_{IV,II_1} \cr
\hline
{\rm Isom} &
\begin{array}{l}
\nearrow {\rm Isom} _{\,V_1, II_2} \cr
\searrow {\rm Isom}_{I} \end{array}
\Bigl\}
\rightarrow
{\rm Isom}_{\, IV, II_1} \cr
\hline
\end{array} $$
\caption{\label{fig-pq}Interconnections among the algebras generated by two
spacelike 1-forms. See Fig.~\ref{fig-lm} for its interpretation, and
text and table~\ref{tb-pq} for details.}
\end{figure}
Although the geometrical structure of both cases is similar,
the Lie algebras strongly depend on their local
character, see also Fig.~\ref{fig-pq}.
This dependence is present in conformal symmetries too, when
dealing with its ``isometrization'', see \cite{bilyalov}--\cite{hst}, where
the signature of the metric is crucial.
For the sake of brevity, we shall only show the results of the
more special situations. For instance, cases $ I_a $, $ I_b $, $ II_1 $,
$ IV_a $, $ IV_b $ and $ V_1 $ can easily be recovered from
Prop.~\ref{prop-pq}
(see also Fig.~\ref{fig-pq}). For the case $ III $, one has the
following
result
\bpr
\label{prop-spq}
{\it No} groups proportional to $ (\spq) $ exist.
\epr
This case sets $ h = f = 0 $ and $ c \neq 0 $. However it turns out that
condition~\rf{eq20} implies $ c = 0 $ and, therefore, one gets a
contradiction.
\N

Next, one finds case $ II_2 $, defined by $ h =  f $, $ c = 0 $.
In this case, contrary to the case $ II_1 $, $ \gamma_1 $ remains free.
Moreover,
the familial relation is linear, so that it is a well-defined case of a
metric
motion. In this case $ {q} = 2 h \, (\pp + \qq) $. Therefore it may be
interpreted as a semi-conformal motion, restricted to the spacelike
part of the metric tensor. Its properties are
$$\displaylines{
\cases{\lh' = \alpha_0 \lh, \quad \mh' = -\alpha_0 \mh, \cr
\ph' =h \ph + \gamma_1 \qh, \quad \qh' =h \qh  - \gamma_1 \ph ,}
\quad {\tilde h} = h' + 2 h^2, \quad \eta' = 2 h \, \eta.
} $$

For the case $ V_2 $, as $ \xiv $ must form a vector space,
the weights must be related with each other at most linearly.
Therefore, they must satisfy
$ h = a^2 f , c = \pm 2 a f $, where $ a $ is a constant
under the action of the motion, cf. Case $ V_2 $ of Sect.~\ref{ss-lglm}.

Following similar steps as in that case, one obtains
\beq
{q} = \Bigl( { 2 f \over a^2 + 1}\Bigr)
\rh_{\pm} \otimes \rh_{\pm} , \nnb
\eeq
with $ \rh_{\pm} \cdot \rh_{\pm} =  1 $,
$  \rh_{\pm} \cdot \rh_{\mp} = 0 $.  Therefore, we
conclude that, in the $ \ph $--$ \qh $ algebras, the case $ V_2 $
is {\it equivalent} to the case $ I_a $, or either $ I_b $.

The Lie algebras generated by a timelike and a spacelike 1-form field, $ \uh
$, $ \nh $ respectively, are
given in App.~\ref{app-c}. Here we just show the summary.

\begin{table}
$$ \begin{array}{||c||l||c||}
\hline
{q} & {\rm Case} & \!\hbox{Extra freedoms} \cr
\hline
\hline
2h\, \uu & I_a & 3 \cr
\hline
2f \, \nn & Ib & 3 \cr
\hline
2h \, \uu + 2f \, \nn &
\begin{array}{l} II_1: h \neq -f \cr II_2: h = -f  \end{array} &
\begin{array}{l} II_1: 1 \cr II_2:2 \end{array} \cr
\hline
c \, (\sun) & III & \not\!\exists \cr
\hline
2 h\, \uu + 2c \, (\sun) & IV_a & 1 \cr
\hline
2 f \, \nn + 2c \, (\sun) & IV_b & 1 \cr
\hline
\begin{array}{l} 2h \, \uu + 2f \, \nn \cr + c \,  (\sun) \end{array} &
\begin{array}{l} V_1: c^2 \ne 4 h f \cr V_2: h \!= \!a^2 f, c\! =\! \pm 2 a
f
\end{array} &
\begin{array}{l} V_1: 2 \cr V_2: 3,4 \end{array} \cr
\hline
\end{array} $$
\caption{\label{tb-un}Generalized motions generated
by $ (\uh, \nh ) $.
The non-existence of Lie algebras for the case $ III $
is a consequence of the stability property. $ a $ is
a constant under the action of the motion $ V_2 $. The latter is
equivalent  to cases $ I_a $, $ I_b $ or I, of table~\ref{tb-lm},
depending on
whether $ a > 1$, $ a < 1 $ or $ a=1 $ (see the App.~\ref{app-d}.}
\end{table}
\begin{figure}
$$ \begin{array}{|c|l|}
\hline
V_1 &\rightarrow II_2 \cr
\hline
(IV_a)_{\, V_1}  = IV_a & \cr
\hline
(IV_b)_{\, V_1} & \rightarrow IV_b \cr
\hline
(II_1)_{\, V_1} & \rightarrow II_1 \cr
\hline
I_a & \rightarrow (I_a)_{\, IV_a,II_1} \cr
\hline
I_b & \rightarrow (I_b)_{\, IV_b,II_1} \cr
\hline
{\rm Isom} &
\begin{array}{l}
\nearrow {\rm Isom} _{\,V_1, II_2} \cr
\rightarrow {\rm Isom}_{\, I_a} \cr
\searrow {\rm Isom}_{\,I_b} \end{array}
\Biggl\}
\rightarrow
{\rm Isom}_{\, IV_a, IV_b,II_1} \cr
\hline
\end{array} $$
\caption{\label{fig-un}Interconnections among the algebras generated by a
spacelike and a timelike 1-forms. See Fig.~\ref{fig-lm} for their
interpretation, and table~\ref{tb-un} for details.}
\end{figure}
\subsection{The addition of a conformal motion to a metric motion}
\label{ss-acg}

In general, once a new metric motion is defined,
one could further add the conformal group. In the case
of \ks symmetries this would mean to consider the problem
$$ \lie g = 2 \phi g + 2h \ll . $$
For instance, the algebraic classification with respect to $ g $
does not change.
This may suggest considering the addition of conformal
motions as a completion of any metric motion.
However, its differential role is completely changed
by its addition, and both problems, the non-conformal and the
conformal one, deserve individual attention (compare isometric
and conformal motions). Nevertheless, in our case, both problems are solved
separately, see e.g., \cite{yano} for the conformal
symmetry. For the \ks case,
the solution is not so simple. The system is mainly closed and general
solutions are to be found as shown in Sect.~\ref{s-ksm}.
For these ``normal'' situations,
the conformal \ks problem might be solved in general as well.
The point is to benefit from the linearity that gives the infinitesimal
study of any metric group.

Conformal \ks relations have been recently
studied in the literature, see e.g., \cite{kramer,mh}. In this
last reference, the authors show that any static spherically
symmetric spacetime can be related to flat spacetime locally
by means of a conformal \ks relation.
They may also be useful in order to describe some
field theories in General Relativity.\footnote{A.Ya. Burinskii, private
communication.}
The translation of these relations into a
problem of metric symmetries, as it was the case for \ks
symmetries alone, benefits from a major simplification:
only first-order terms are needed. We have started to
work on this idea. Yet, at
the moment, the expressions are still under study. But the
main issue deserves some attention because it can also be
extended to any other compound problem of symmetries.
\section{Some remarks concerning possible ways of future research}
\label{ss-rpa}
It is apparent from earlier works that no single, and simple, path can solve
most of current
problems in the field of symmetries and in the search of useful solutions to
Einstein's field equations. Hence, in this section we point out
some lines that we think will help towards a fuller comprehension of
generalized metric symmetries.

The following hints try to ponder the
simplicity of the problem and the reaching of the expected solutions. They
follow the order of former sections. Except the last two, the rest are
already under
consideration with other authors, or by myself.

\ni -- First, a study of \ks motions for the cases of non-geodesic $ \lh $ and
geodesic $ \lh $ with $ \Delta \neq 0 $ through their connection with
isometries.
And, second, find the integrals of motion associated with these motions.
(See Sects.~\ref{ss-ngk},~\ref{ss-gldn0}).

\ni -- To try to obtain a complete solution for \ks motions with geodesic $ \lh
$ and $ \Delta = 0 $, at least in most relevant spacetimes. For this goal
the Newman-Penrose formalism may be helpful.
(See Sect.~\ref{sss-sld0}).

\ni -- To try to find a general solution of \ks motions in flat spacetime.
This would yield a complete knowledge of \ks relations that do not yield new
spacetimes. (See App.~\ref{app-d}).

\ni -- To investigate the relation of generalized metric motions with
collineat\-ions. This path opens a different viewpoint to dealing with
infinite dimensional Lie algebras in a Riemannian manifold. This can be
started from the expressions of $ \lie \Gamma^{\alpha}_{\beta \gamma} $ and
$ \lie R^{\alpha}_{\; \beta \gamma \delta} $ in terms of $ q $ ($\equiv \lie
g $) ---see e.g., App.~\ref{app-a} for the \ks case--- and using the works
referred to in Sect.~\ref{intro}.

\ni -- To address a detailed resolution of other generalized metric motions.
The candidates might correspond either to a pure mathematical interest
---e.g., in connection with first and second before---, to a geometrical
interest, such as conformal or \ks motions are, or to a physical aim, such
as those which arise from Pleba\'{n}ski and Schild's work, or Bonanos' work.

In any case, we would like to mention three other points. First, the study
of
the conditions for the existence of proper vector fields.
Second, its complement. In particular, the
issue of the set of isometries that are compatible with a
generalized metric motion
is perhaps of major interest (for the case of \ks
motions some results will be reported elsewhere). And, finally, the study of
the conditions of maximum integrability of the chosen metric motion and the
structure of the associated Lie algebras.

The case of conformal \ks motions seems to be a good proposal fairly
fulfilling previous criteria.

\ni -- To start an extension of metric motions
to subspaces of a greater
dimension than one. One could follow, for instance, the idea put in
\cite{bonanos2}. In that work the author considered the possibility of
introducing a symmetric 2-covariant
tensor, $ T $, having the algebraical properties of an electromagnetic
energy momentum tensor ---i.e., $ T^{\lambda}_{\lambda} = 0 $, $ T_{\alpha
\lambda} T^{\lambda}_{\beta} = {\sigma}^2 g_{\ab} $. However, no solutions
were obtained. Metric motions may play an important role
in analysing {\em Ans\"atze} with a
geometrical basis. Therefore, combining the author's approach with the ideas
presented here about the role of metric motions would
help towards a better knowledge of that {\em Ansatz}.

-- Finally, to 
develop other possible lines of physical applications. (To that end,
see e.g.,~\cite{pakss}).
\section{Summary and conclusions}
\label{s-con}

Let us summarise very briefly the former sections. Along the preceding sections we have shown mainly two
facts (more details were given in Sect.~\ref{intro} and~\ref{ss-sksm}). Firstly, with
regard \ks motions, we have proven that their notion as metric motion is
meaningful and that their existence and properties
depend on the kinematical properties of the deformation direction $ \lh $
and also that they are generically linked with some subset of the isometries
of a ---generally--- different spacetime.

Second, we have shown that the notion of other metric motions different to
isometries or conformal motions is meaningful and contains a much richer
structure than their predecessors. Despite the fact that any choice is
partly subjective and that a good number of questions still remain to be solved (recall former section) the wealth of results obtained throughout former
sections ---both in connection with isometries and conformal motions and
with new properties--- prove that we were on a right track. In particular,
we remark that the associated Lie algebras may be of an infinite dimensional
character and this may happen in any dimension of the manifold. This result
is absent in isometries and only
holds for conformal motions in 2-dimensional Riemannian manifolds.

Finally, the principal role assigned to geometry in our framework will
certainly help towards the decipheration of old and new physical
applications of metric motions, as to finding new --- physically
interesting--- solutions to Einstein's equations, or in addressing some
basic questions on the whole fabric of motions, i.e., including
collineations.

Overall, an active issue that may stimulate the interconnection and
feedback between some
general mathematical issues of differential
geometry and their applicability in
gravitational physics.

\section*{Acknowledgements}
The author wishes to thank Profs. Bartolom\'e Coll, emilio Elizalde and Graham S. Hall for several useful discussions. he also acknowledges valuable discusiions or comments with Drs. Jaume Carot, Jos\'e Alberto Lobo, diego Pav\'on and Prof. Jos\'e Mar\'\i a M. Senovilla, as well as some comments from anonymous referees and Prof. hans-J\"urgen Schmidt.

\appendix
\section{Miscellaneous of Kerr-Schild-like formulae}
\label{app-a}

\proclaim Lie derivatives of the Levi-Civita connection, the Riemannian
and Ricci tensor and of the scalar curvature.

\ni To begin with,
$$ \lie g_{\ab} = 2 h \, \ell_{\alpha} \, \ell_{\beta},
\quad \lie g^{\ab} = - 2 h \, \ell^{\alpha} \, \ell^{\beta}, $$
where we have used $ \lie g^{\ab} = - g^{\alpha \sigma} g^{\beta \mu} \lie
g_{\sigma \mu} $.

In any coordinate basis
\beq
\label{app-lie-Gamma} 2 \lie \Gamma^{\alpha}_{\beta \gamma} &=
g^{\alpha \lambda} [\nabla_{\beta} \lie g_{\lambda \gamma} + \nabla_{\gamma}
\lie g_{\lambda \beta} - \nabla_{\lambda} \lie g_{\beta \gamma}] \cr
\label{lie-gamma-ks}
&= \nabla_{\beta} (h \ell^{\alpha} \ell_{\gamma} ) + \nabla_{\gamma} (h
\ell^{\alpha} \ell_{\beta} ) - \nabla^{\alpha} (h \ell_{\beta} \ell_{\gamma}
) .
\eeq
This last expression is easily generalized to any other case of metric motion.
Another useful expression ---which will be used in App.~\ref{app-b}--- is
\beq
\label{eq-nor}
\nabla_{\beta} \nabla_{\gamma} \xi^{\alpha} =
R^{\alpha}_{\ \gamma \beta \lambda} \xi^{\lambda} + \lie
\Gamma^{\alpha}_{\beta \gamma} .
\eeq

On the other hand,
\beq
\label{lie-rie-gamma}
\lie R^{\alpha}_{\; \beta \gamma \delta} & = &
2 \nabla_{[\gamma} \lie
\Gamma^{\alpha}_{\delta]\beta} \cr
& = &
h (R^{\alpha}_{\; \lambda
\gamma \delta}\, \ell^{\lambda} \,\ell_{\,\beta} -
R^{\lambda}_{\; \beta
\gamma \delta} \, \ell^{\alpha} \, \ell_{\,\lambda}) +
\nabla_{\gamma}
\nabla_{\beta} ( h \ell^{\alpha} \ell_{\, \delta}) - \nabla_{\delta}
\nabla_{\beta} (h \ell^{\alpha} \ell_{\, \gamma}) \cr
\label{lie-rie-ks}& & + \nabla_{\delta} \nabla^{\alpha} ( h \ell_{\beta}
\ell_{\gamma}) - \nabla_{\gamma} \nabla^{\alpha}
(h \ell_{\beta} \ell_{\delta}) ,
\eeq
where we have used Ricci identities for $ h \ell^{\alpha} \ell_{\beta} $.

Contracting indices $ \alpha $ and $ \gamma $ we get (after renaming $ \beta
$ and $ \delta $)
\beq
\lie R_{\ab}& = & (Dh_{\alpha}) \ell_{\beta} + (Dh_{\beta}) \ell_{\alpha} +
h_{\alpha}
L^{\sigma}_{\ \beta \sigma} + h_{\beta} L^{\sigma}_{\ \alpha
\sigma} \cr
&& + h_{\sigma} \bigl( L^{\sigma}_{\ \alpha \beta} +
L^{\sigma}_{\ \beta \alpha} - 2 L_{\ab}^{\ \, \ \sigma} \bigr)
+ h \bigl( L^{\sigma}_{\ \alpha \beta \sigma} + L^{\sigma}_{\
\beta \alpha \sigma} - L^{\quad \ \sigma}_{\alpha \beta \sigma} \bigr) \cr
&& - g^{\sigma \mu} \bigl( \nabla_{\sigma} h_{\mu} \bigr)
\ell_{\alpha}\ell_{\beta},
\eeq
where $ h_{\alpha} \equiv \nabla_{\alpha} h $, $ L^{\alpha}_{\ \beta
\gamma} \equiv \nabla_{\gamma} \bigl( \ell^{\alpha} \ell_{\beta} \bigr) $
and $ L^{\alpha}_{\ \beta \gamma \delta} \equiv \nabla_{\delta}
\nabla_{\gamma} \bigl( \ell^{\alpha} \ell_{\beta} \bigr) $ 
(see notation
in Sect.~\ref{intro} for the rest). We have also used that 
$ R_{\sigma \delta} \ell^{\sigma} \ell_{\beta} 
- R^{\lambda}_{\ \beta \sigma \delta} \ell^{\sigma} \ell_{\lambda} + L^{\sigma}_{\ \beta \sigma \delta} 
= L^{\sigma}_{\ \beta \delta \sigma} $ in order to express the result in a manifestly symmetric form. The quantities 
$ L^{\sigma}_{\ \alpha \sigma} $, 
$ L^{\sigma}_{\ \alpha \beta \sigma} $
can be expressed 
in terms of some 
kinematical quantities of $ \lh $. They read: $ L^{\sigma}_{\ \alpha \sigma} =  2 \theta \ell_{\alpha} + a_{\alpha} $, $ L^{\sigma}_{\ \alpha \beta \sigma} = 
( 2 \nabla_{\beta} \theta + R_{\lambda \beta} \ell^{\lambda}) \ell_{\alpha} + (\nabla_{\beta} \ell^{\sigma})(\nabla_{\sigma} \ell_{\alpha}) + 2 \theta \nabla_{\beta} \ell_{\alpha} + D(\nabla_{\beta} \ell_{\alpha}) $.

The Lie derivative of the scalar curvature with respect to a KSVF is $ \lie
R = \lie (g^{\lambda \mu} R_{\lambda \mu} ) = - 2
h\, \ell^{\lambda} \ell^{\mu} R_{\lambda \mu} + g^{\lambda \mu} \lie
R_{\lambda \mu} = - 2 h {\bar R} + g^{\lambda \mu} \lie R_{\lambda \mu} $.
Thus,
\beq
\lie R = 2 \bigl\{ DDh + 4 \theta \,Dh + h [ 2(D\theta + 2 \theta^2 ) +
\nabla_{\mu} a^{\mu} - {\bar R} ] + a^{\mu} h_{\mu} \bigr\}. \nnb
\eeq
Whence, the Lie derivative of the Einstein's tensor, $ G_{\ab} \equiv
R_{\ab} - (1/2) g_{\ab} R $, with respect to a KSVF can easily be
obtained ---we do not need it here.

Let us finally recall that in the case of a geodesic $ \lh $ other useful
expressions are already given in~\rf{gr}--\rf{grie}.

\proclaim Commutation of covariant and Lie derivatives.

\ni Let $ T^{\alpha_1 \cdots \alpha_p}_{\beta_1 \cdots \beta_q} $ be the
components of a $ p $-contravariant, $ q $-covariant tensor field. A
standard computation gives
$$ \displaylines{ [\lie, \nabla_{\gamma}] \, T^{\alpha_1 \cdots
\alpha_p}_{\beta_1 \cdots \beta_q} =
  \sum_{i=1}^{i=p}
(\lie \Gamma_{\gamma\lambda}^{\alpha_i})
T^{\alpha_1 \cdots \alpha_{i-1} \lambda \, \alpha_{i+1} \cdots
\alpha_p}_{\beta_1
\cdots \beta_q} \hfill{} \cr \hfill{}- \sum_{i=1}^{i=q}
(\lie \Gamma_{\gamma \beta_i}^{\lambda})
T^{\alpha_1 \cdots \alpha_p}_{\beta_1 \cdots \beta_{i-1}
\, \lambda \,\beta_{i+1} \cdots \beta_{q} }. }$$
For the case of the null deformation direction $ \lh $ we
readily get
\beq
\label{app-1}
[\lie,\nabla_{\gamma}] \ell^{\alpha} = - D(h \, \ell^{\alpha}
\ell_{\gamma}), \quad [\lie,\nabla_{\gamma}] \ell_{\alpha} = - D(h \,
\ell_{\alpha} \ell_{\gamma}),
\eeq
where we have used~\rf{lie-gamma-ks}.

\proclaim Lie derivatives of kinematic quantities of $ \lh $.

\ni First of all, from Eqs.~\rf{eq-h},~\rf{eq-sta},
\beq
\label{app-2}
\lie \lv = m \, \lv
\eeq
because $ \lie \ell^{\alpha} = (\lie g^{\alpha \sigma}) \ell_{\sigma} +
g^{\alpha \sigma} \lie \ell_{\sigma} = m \ell^{\alpha} $.

In the sequel we will freely use previous results in the proofs. We then
obtain
\begin{itemize}
\item[i.-] $ \label{app-th}\lie \theta  =  m \, \theta + {1 \over 2} Dm $.
\item[ii.-] $ \label{app-dth}\lie \, D\theta  =  2m \, D\theta + (Dm)\,
\theta + {1 \over 2} DDm $.
\end{itemize}
\pr In~i we have $ 2 \lie \theta = \lie \nabla_{\lambda} \ell^{\lambda} =
\nabla_{\lambda} \lie \ell^{\lambda} = \nabla_{\lambda} (m \ell^{\lambda}) =
2m \, \theta + Dm $. For~ii we have $ \lie \, D\theta = \lie (
\ell^{\lambda} \nabla_{\lambda} \theta) = (\lie
\ell^{\lambda}) \nabla_{\lambda} \theta + \ell^{\lambda}\lie
\nabla_{\lambda} \theta = m \, D\theta + \ell^{\lambda} \nabla_{\lambda}
\lie \theta = m \, D \theta + D[m\theta + (1/2)Dm] = 2m \, D \theta + (Dm)
\theta + (1/2) DDm $.\N

In the case of a geodesic $ \lh $ ---$ D \lv = M \lv $--- one also has
\begin{itemize}
\item[iii.-] $ \label{app-br}\lie {\bar R} = 2 m {\bar R}$.
\item[iv.-] $ \label{app-en}
\lie \ell^{\mu}\nabla_{\sigma} \nabla^{\sigma} \ell_{\mu} = 2m
\ell^{\mu}\nabla_{\sigma} \nabla^{\sigma} \ell_{\mu}$.
\item[v.-] $ \label{app-m}\lie M = m \, M + Dm $.
\item[vi.-] $ \label{app-dm}\lie DM = 2m \, DM + (Dm) M + DDm $.
\end{itemize}
\pr In~iii one has $ \lie (R_{\lm} \ell^{\lambda} \ell{\mu} ) =
\ell^{\lambda}\ell^{\mu} \lie R_{\lm}
+ 2m {\bar R} = 2m {\bar R} $, because 
$ \ell^{\lambda}\ell^{\mu} 
\lie R_{\lm}= - 2 h a^2 $ and vanishes for a geodesic $ \lh $. For~iv it is a bit longer. We have $ \lie (
\ell^{\mu}\nabla_{\sigma} \nabla^{\sigma} \ell_{\mu}) = m\,
\ell^{\mu}\nabla_{\sigma} \nabla^{\sigma} \ell_{\mu} + \ell^{\mu} \lie (
\nabla_{\sigma} \nabla^{\sigma} \ell_{\mu}) = m 
\, \ell^{\mu}\nabla_{\sigma}
\nabla^{\sigma} \ell_{\mu} + \ell^{\mu} \nabla_{\sigma} (\lie \nabla_{\sigma}
\ell_{\mu} )+ \ell^{\mu}(\lie \Gamma^{\sigma}_{\lambda \sigma})\nabla^{\lambda}
\ell_{\mu} - \ell^{\mu}(\lie\Gamma^{\lambda}_{\mu \sigma})\nabla^{\sigma}
\ell_{\lambda} $; the two latter terms cancel. On the other hand, $ \lie
\nabla^{\sigma} \ell_{\mu} = - 2 M \, h \, \ell^{\sigma} \ell_{\mu} +
g^{\sigma \rho} \lie \nabla_{\rho} \ell_{\mu} = - 2 M \, h \, \ell^{\sigma}
\ell_{\mu} + \nabla^{\sigma}(m\ell_{\mu}) + (Dh)
\ell^{\sigma}\ell_{\mu} + 2
M\, h \, \ell^{\sigma} \ell_{\mu} = \nabla^{\sigma}(m\ell_{\mu}) + (Dh)
\ell^{\sigma}\ell_{\mu} $. Therefore $ \ell^{\mu}
\nabla_{\sigma} \lie \nabla^{\sigma} \ell_{\mu} = m \,
\ell^{\mu}\nabla_{\sigma} \nabla^{\sigma} \ell_{\mu} $, and thus one gets
the claimed result.
For~v we have
$ \lie ( M \ell^{\alpha} ) = \lie (D \ell^{\alpha}) = 
\lie (\ell^{\mu} \nabla_{\mu}
\ell^{\alpha} ) = m \, M \, \ell^{\alpha} + D(m\ell^{\alpha}) - \ell^{\mu} D(h
\ell^{\alpha} \ell_{\mu}) = (2m\, M + Dm) \ell^{\alpha} $.
On the other hand, $ \lie (M \ell^{\alpha}) = (\lie M)\ell^{\alpha} + mM\ell^{\alpha} $ and, hence, one gets v.
For~vi one can follow the proof of ii. \N

This list can easily be completed for other quantities and for a
non-geodesic $ \lh $. They were not used in this work and therefore we do
not display them here.
\section{First steps towards the integrability equations of a metric motion}
\label{app-b}

\proclaim The general case.

\ni First of all one needs to write the system of
Eqs.~\rf{eq8}--\rf{eq9} in a normal form {\it for} the basic
unknowns $ \xiv $
and its derivatives, at least. We will assume a coordinate basis.
The first result is that Eqs.~\rf{eq8}--\rf{eq9} are equivalent
to the partial differential system ($ q $, $ m $ are used instead of
$ \qx $, $ m_{\xi} $, respectively,
to clarify the notation)
\beq
\label{eq-xi}
\cases{\nabla_{\alpha} \xi_{\beta} =
\xi_{\alpha \beta}, \quad \xi_{\ab} +
\xi_{\beta \alpha} = q_{\ab} = (q_{\ol})
\Theta^{\Omega}_{\,\alpha}  \Theta^{\Lambda}_{\,\beta}, \cr
\xi^{\lambda}\nabla_{\lambda} \Theta^{\Omega}_{\,\alpha} +
\xi^{\; \lambda}_{\alpha} \Theta^{\Omega}_{\,\lambda}  =
m^{\Omega}_{\Lambda} \Theta^{\Lambda}_{\, \alpha}, \cr
\begin{array}{lrl}
\nabla_{\alpha} \xi_{\beta \gamma} & = &
R_{\gamma \beta \alpha \lambda} \xi^{\lambda}
+{1 \over 2} (\nabla_{\alpha} q_{\gamma \beta}+
\nabla_{\beta} q_{\gamma \alpha}-\nabla_{\gamma}
q_{\ab}) .\cr
\end{array}}
\eeq
In the computation of $ \nabla_{\alpha} \xi^{. \gamma}_{\beta} $
we have used Eq.~\rf{app-lie-Gamma} and Eq.~\rf{eq-nor}.
Whence, a standard ---tough cumbersome---
computation shows that (we put $ A_{<\alpha \gamma>}
\equiv  ( 1 / 2 ) ( A_{\alpha_1\alpha_2\gamma_1\gamma_2} +
A_{\gamma_1\gamma_2\alpha_1\alpha_2} ) $)
\bpr
\label{prop-intcond}
Ricci identities applied to a metric motion can be expressed as
\beq
\label{first-int2}
\xi^{\lambda} \nabla_{\lambda}
R_{\alpha_1 \alpha_2 \gamma_1 \gamma_2}& +
& 4 \xi_{[\rho \lambda]} (\delta^{\rho}_{[\alpha_1}
R^{\lambda}_{\; \alpha_2] \gamma_1 \gamma_2})_{<\alpha \gamma>}
= \cr
& & - q_{\lambda [\gamma_1}
R^{\lambda}_{\; \gamma_2]\alpha_1 \alpha_2}
- 2 \nabla_{[\gamma_1}\nabla_{[\alpha_1}
q_{\alpha_2]\gamma_2]} ,
\eeq
\beq
\label{sec-int2}
\xi^{\lambda} \nabla_{\lambda}\nabla_{\alpha_3}
R_{\alpha_1 \alpha_2 \gamma_1 \gamma_2} +\!
\xi_{[\rho \lambda]} \Bigl[ 4 \nabla_{\alpha_3}
(\delta^{\rho}_{[\alpha_1}
R^{\lambda}_{\; \alpha_2] \gamma_1 \gamma_2})_{<\alpha \gamma>}
\! \!+ \!\delta^{\rho}_{\alpha_3} \nabla^{\lambda}
R_{\alpha_1 \alpha_2 \gamma_1 \gamma_2}\Bigr] = \hfill{}\cr
- {\textstyle {1 \over 2}} q_{\, \alpha_3}^{\lambda}\nabla_{\lambda}
R_{\alpha_1 \alpha_2 \gamma_1 \gamma_2} +
2 \Bigl[\Bigl(\nabla_{\lambda}
q_{\alpha_3 [\alpha_1}-\nabla_{[\alpha_1,} q_{\lambda \alpha_3} \Bigr)
R^{\lambda}_{\; \alpha_2]\gamma_1 \gamma_2}\Bigr]_{<\alpha \gamma>}
- \cr \nabla_{\alpha_3}( q_{\lambda [\gamma_1}
R^{\lambda}_{\; \gamma_2]\alpha_1 \alpha_2}) -
2 \nabla_{\alpha_3} \nabla_{[\gamma_1}\nabla_{[\alpha_1}
q_{\alpha_2]\gamma_2]} ,
\eeq
$$ \ldots $$
\beq
\label{stab2}
\xi^{\lambda}\nabla_{\lambda} \Theta^{\Omega}_{\, \alpha} +
\Theta^{\Omega \lambda}\xi_{[\alpha \lambda]} =
- {\textstyle {1 \over 2}}
q^{\Omega}_{\, \alpha} +
m^{\Omega}_{\, \Lambda} \Theta^{\Lambda}_{\, \alpha},
\eeq
\beq
\label{der-stab2}
\xi^{\lambda}\nabla_{\lambda} \nabla_{\alpha}
\Theta^{\Omega}_{\, \beta} + \xi_{[\lambda \sigma]}
(\delta^{\lambda}_{\alpha}\nabla^{\sigma}\Theta^{\Omega}_{\, \beta} +
\delta^{\lambda}_{\, \beta}\nabla_{\alpha}\Theta^{\Omega \sigma}) = \cr
(\nabla_{\alpha} m^{\Omega}_{\, \Lambda}) \Theta^{\Lambda}_{\, \beta} +
{\textstyle {1 \over 2}}(\nabla^{\lambda} q_{\ab}-
\nabla_{\beta} q^{\lambda}_{\, \alpha})\Theta^{\Omega}_{\, \lambda} +
q^{\Omega}_{\, \Lambda}\nabla_{\alpha} \Theta^{\Lambda}_{\, \beta} -
{\textstyle {1 \over 2}} q_{\beta \lambda} \nabla_{\alpha}
\Theta^{\Omega \lambda} -
{\textstyle {1 \over 2}} \nabla_{\alpha} q^{\Omega}_{\, \beta},
\eeq
$$ \ldots $$
and the Ricci identities for the functions $ m^{\Omega}_{\, \Lambda} $, $
q_{\ol} $ and their derivatives.
\epr
\pr We will only give a sufficient sketch of it.
The Ricci identities applied to $ \xi_{\alpha} $ are
identically satisfied by virtue of Eqs.~\rf{eq-xi}. The following
are the Ricci identities for $ \xi^{\alpha}_{\beta} $. A standard
computation (see e.g., \cite{yano})
leads to the conclusion that
these are in fact collected in the form of Eq.~\rf{lie-rie-gamma}. Using
Eq.~\rf{app-lie-Gamma} one gets Eq.~\rf{lie-rie-ks}
for a generic $ q $: 
\beq
\label{lie-rie}
\lie R^{\alpha}_{\; \beta \gamma \delta} & = &
R^{\alpha}_{\; \lambda
\gamma \delta}\, q^{\lambda}_{\,\beta} +
R^{\lambda}_{\; \beta
\gamma \delta} \, q^{\alpha}_{\,\lambda} +
\bigl( \nabla_{\gamma}
\nabla_{\beta} q^{\alpha}_{\, \delta} - \nabla_{\delta}
\nabla_{\beta} q^{\alpha}_{\, \gamma} \bigr) \cr
& & + \bigl( \nabla_{\delta} \nabla^{\alpha} q_{\beta \gamma} -
\nabla_{\gamma} \nabla^{\alpha} q_{\beta \delta} \bigr).
\eeq
Of course, as emphasized elsewhere, some knowledge on $
q $ must be given in order to have a well-defined problem. It is
worth writing them, as long as possible, in terms of the independent
variables $ \xi_{\alpha} $, $ \xi_{[\ab]} $. Substituting the constraint
given by Eqs.~\rf{eq8} into
Eqs.~\rf{lie-rie}  one gets Eqs.~\rf{first-int2}. The following conditions
are Ricci identities applied to $ \nabla_{\nu} \xi^{\, \alpha}_{\beta} $.
Once again, a standard analysis,
shows that these
can be indeed  written calculating
$ \lie \nabla_{\gamma} R^{\alpha}_{\, \beta \gamma \delta} $. In fact
its calculation may be carried out more easily via the commutation
identity
\beq
\label{lie-nabla}
[\lie, \nabla_{\nu}] R^{\alpha}_{\; \beta \gamma \delta} = \cr
(\lie \Gamma^{\alpha}_{\nu \lambda}) R^{\lambda}_{\; \beta
\gamma \delta} - (\lie \Gamma^{\lambda}_{\nu
\beta})R^{\alpha}_{\; \lambda \gamma \delta} - (\lie
\Gamma^{\lambda}_{\nu \gamma})R^{\alpha}_{\; \beta \lambda
\delta} - (\lie \Gamma^{\lambda}_{\nu \delta})R^{\alpha}_{\; \beta
\gamma \lambda}.
\eeq
It is worth writing the result
in terms of $ \xi_{\alpha} $, $ \xi_{[\ab]} $ only, and this may
be performed as in the previous case. The result is~\rf{sec-int2}.

This procedure has to be continued to any order (notice that only
$ \xi_{\alpha} $ and $ \xi_{[\ab]} $ appear at each step). This is a usual
presentation in the case of isometric or conformal motions, of course,
setting $ q $ in accordance. The
difference now lies in the fact that the system may be open
---except in the case of isometric or conformal motions where
the system is always normal, see e.g.,~\cite{yano}. Thus one
can only include the Ricci identities for $ m^{\Omega}_{\Lambda} $, $
q_{\Omega \Lambda} $ and their
derivatives in a direct way.

Secondly, Eq.~\rf{eq-sta} and its derivatives add another set of conditions
for any other case which differs from isometric or conformal
motions. In the previous case, Eq.~\rf{eq-h} was directly substituted into
the system, yet this is not possible for the stability conditions,
Eqs.~\rf{eq-sta} or their
derivatives, because of their expression (neither
in the new case of \ks motions) and they
must be added directly. Again, for any derivative, one
must use the commutation identity~\rf{lie-nabla} (adapted to
each case), in order to write them in terms of $ \xi_{\alpha} $
and $ \xi_{[\ab]} $ only. These are Eqs.~\rf{stab2},~\rf{der-stab2}, \ldots
$\, $ Finally, in the expressions lower case Greek indices are raised or
lowered
using the tensor metric $ g_{\ab}$, and upper case ones with the
aid of $ g^{\Omega \Lambda} \equiv \ofo \cdot \ofl $ and its
inverse. \N

\proclaim The \ks case.

\ni
This case is recovered from
the general expression setting $ q = 2 h \, \ll $ although some
simplifications appear.
Eqs.~\rf{eq-xi} turn into
\beq
\label{ks-normal}
\cases{\xi_{(\ab)} = h \ell_{\alpha} \ell_{\beta}, \quad
\xi^{\rho}\nabla_{\rho}
\ell_{\alpha} + \ell^{\rho}\xi_{\alpha \rho} =m \ell_{\alpha}, \cr
\nabla_{\alpha} \xi_{\beta} = \xi_{\alpha \beta}, \quad \nabla_{\alpha}
h =
h_{\alpha},\cr
\!\!\!$$\begin{array}{lrcr}
\nabla_{\alpha} \xi_{\beta \gamma}& = &
R_{\gamma \beta \alpha \lambda} \xi^{\lambda} + h_{\beta}
\ell_{\gamma}\ell_{\alpha} + h_{\alpha} \ell_{\gamma}\ell_{\beta}
- h_{\gamma} \ell_{\alpha}\ell_{\beta} \cr
&& + h[\nabla_{\alpha} (\ell_{\beta}\ell_{\gamma}) +
\nabla_{\beta}(\ell_{\alpha}\ell_{\gamma}) -
\nabla_{\gamma}(\ell_{\alpha}\ell_{\beta})] .\cr
\end{array}}
\eeq

Therefore, it is clear that it suffices to consider $ \lh $ as the only
relevant element of the cobasis. Whence, using the same notation as in
Prop.~\ref{prop-intcond} we get:
\bpr
\label{icks}
Ricci identities applied to a \ks motion are
\beq
\label{first-int}
\xi^{\sigma} \nabla_{\sigma}
R_{\alpha_1 \alpha_2 \gamma_1 \gamma_2}& +
&4 \xi_{[\rho \lambda]}(\delta^{\rho}_{[\alpha_1}
R^{\lambda}_{\; \alpha_2] \gamma_1 \gamma_2})_{<\alpha \gamma>}
= \cr
& & - q_{\lambda [\gamma_1}
R^{\lambda}_{\; \gamma_2]\alpha_1 \alpha_2}
- 2 \nabla_{[\gamma_1}\nabla_{[\alpha_1}
q_{\alpha_2]\gamma_2]} ,
\eeq
\beq
\label{sec-int}
\xi^{\sigma} \nabla_{\sigma}\nabla_{\alpha_3}
R_{\alpha_1 \alpha_2 \gamma_1 \gamma_2} +\!
\xi_{[\rho \lambda]} \Bigl[ 4 \nabla_{\alpha_3}
(\delta^{\rho}_{[\alpha_1}R^{\lambda}_{\; \alpha_2]
\gamma_1 \gamma_2})_{(\alpha \gamma)}\! \!+
\!\delta^{\rho}_{\alpha_3} \nabla^{\lambda}
R_{\alpha_1 \alpha_2 \gamma_1 \gamma_2}\Bigr]
= \hfill{}\cr
- {1 \over 2} q_{\alpha_3}^{\,\lambda}\nabla_{\lambda}
R_{\alpha_1 \alpha_2 \gamma_1 \gamma_2}+
2 \Bigl[\Bigl(\nabla_{\lambda}
q_{\alpha_3 [\alpha_1}-\nabla_{[\alpha_1,}
q_{\lambda \alpha_3} \Bigr) R^{\lambda}
_{\; \alpha_2]\gamma_1 \gamma_2}\Bigr]_{<\alpha \gamma>}
- \cr
\nabla_{\alpha_3}( q_{\lambda [\gamma_1}
R^{\lambda}_{\; \gamma_2]\alpha_1 \alpha_2}) -
4 \nabla_{\alpha_3} \nabla_{[\gamma_1}\nabla_{[\alpha_1}
q_{\alpha_2]\gamma_2]} ,
\eeq
$$ \ldots $$
\beq
\label{stab}
\xi^{\sigma}\nabla_{\sigma} \ell_{\alpha} - \xi_{[\sigma \alpha]}
\ell^{\sigma} = m \, \ell_{\alpha},
\eeq
\beq
\xi^{\sigma}\nabla_{\sigma} \nabla_{\alpha}\ell_{\beta}
+ \xi_{[\lambda \sigma]} (\delta^{\lambda}_{\alpha}
\nabla^{\sigma}\ell_{\beta} +
\delta^{\lambda}_{\beta}\nabla_{\alpha}\ell^{\sigma})
& = &
\!  m \nabla_{\alpha} \ell_{\beta}\nnb \\ 
& & \! + \ell_{\beta}
[ \nabla_{\alpha} m + \! h\,a_{\alpha}\! + \! (Dh) \ell_{\alpha}],
\label{der-stab}
\eeq
$$ \ldots $$
and the Ricci identities for $ m $, $ h $ and their derivatives. \N
\epr
Due to the little closure of the system in general, the above equations may
be viewed
as {\em first steps} towards
the integrability conditions of
a generalized metric motion. 
They may also be useful to study maximum
integrability of a given generalized metric motion.
\section{Metric motions generated by $ \uh $--$ \nh $}
\label{app-c}

Let $ \uh $ and $ \nh $ be two 1-forms satisfying $ \uh \cdot \uh = -1$,
$\nh \cdot \nh = 1 $, $ \uh \cdot \nh = 0 $.

The differential expression of a metric motion
generated by these two ingredients is
\beq
\label{qun}
\lie \bg = 2h \uu + c(\sun) + 2f \nn.
\eeq
In this scheme $ \bg $, $ \uh $, and $ \nh $ may be regarded as data.
On the other hand, $ h $, $ f $, $ c $
are unknown $ C^{\infty} $ functions of the manifold yet to
be determined. And
$ \xiv $ are the infinitesimal generators of the group.
The cobasis will be completed with the addition of two
spacelike 1-forms, $ \ph $ and $ \qh $, satisfying
$ \ph \cdot \ph = \qh \cdot \qh = 1 $, $ \uh \cdot \ph = \uh \cdot \qh
= \nh \cdot \ph = \nh \cdot \qh = \ph \cdot \qh  = 0 $, but
otherwise arbitrary.

We summarize the results of this section as follows ($ A' $ stands for $
\lie A
$)
\bpr[Metric motions generated by $ u $--$ n $]
\label{prop-un}
The conditions in order
to have a generalized metric motion generated by $ \uh $--$ \nh $, i.e.,
Eqs.~\rf{qun}, are
\beq
\label{eq21}
\cases{\uh' = - h \uh + \alpha_0 \nh + \alpha_1 \ph + \alpha_2 \qh,
\quad \nh' = (c + \alpha_0) \uh + f \nh +  \beta_1 \ph + \beta_2 \qh, \cr
\ph' = \alpha_1 \uh - \beta_1 \mh + \gamma_1 \qh,
\quad \qh' = \alpha_2 \uh -  \beta_2 \nh  - \gamma_1 \ph,}
\eeq
with
\beq
\label{eq22}
\cases{ 2 h \alpha_1 + c \beta_1 = 0, \quad 2 f \beta_1 + c \alpha_1 = 0,
\cr
2 h \alpha_2 + c \beta_2 = 0, \quad 2 f \beta_2 + c \alpha_2 = 0,}
\eeq
and
\beq
\label{eq23}
\cases{
{\tilde h} = h' - 2 h^2 + c ( c + \alpha_0) , \cr
{\tilde f} = f' + 2 f^2 +  c \alpha_0, \cr
{\tilde c} = c' + 2 \alpha_0 ( h + f ) + c ( 3 f - h ).}
\eeq
The volume element transforms according to
\beq
\label{vol-un}
\eta' = (f-h) \, \eta .
\eeq
\epr
In these expressions $ \{ \uh $, $ \nh $, $ \qh $, $\ph \} $ is
{\it any} orthonormal cobasis of the manifold and
$ \{\alpha_0 $, $ \alpha_1 $, $ \alpha_2 $,
$ \beta_1 $, $\beta_2 $, $\gamma_1\} $
are $ C^{\infty}$ functions.

\pr The proof follows analogous steps as those of Prop.~\ref{prop-lm}
(now $ \bg = - \uu + \nn + \allowbreak\pp + \allowbreak\qq $). \N

One also has
\bpr
\label{prop-dis-un}Isometries and the cases $ I_a $, $ I_b $, $ II_1 $,
$ IV_a $, $ IV_b $, $ V_2 $ are disconnected from the general
solution, case $ V_1 $.
\epr \N

Again, for the sake of brevity, we shall only display some cases,
see also Fig.~\ref{fig-un} for a summary.
Cases $ I_a $, $ I_b $, $ II_1 $, $ IV_a $,
$ IV_b $, and $ V_1 $ are easily recovered from
Prop.~\ref{prop-un}. Notice, however, that now cases ``$ a $''
and ``$ b $'' are not equivalent because $ \uh $ and $ \nh $ are
timelike and spacelike, respectively. Furthermore, case
$ II_2 $ is very similar to $ II_2 $ of the last section.
Finally, for the case $ III $ one has
\bpr
\label{prop-sun}
{\it No} motions proportional to $ (\sun) $ exist.
\epr \N

And for the case $ V_2 $, as $ \xiv $ must form a vector space, the
weights must satisfy $ h = a^2 f , g = \pm 2 a f $, where $ a $ is
a constant under the action of the group, cf.~case $ V_2 $
Sect.~\ref{ss-lglm}.
Following similar steps as in that case,
one obtains ($ a^2 \neq 1$)
\beq
{q} = \Bigl( { 2 f \over 1- a^2}\Bigr)
\rh_{\pm} \otimes \rh_{\pm}, \nnb
\eeq
with $ \rh_{\pm} \cdot \rh_{\pm} =  \sign(1-a^2) $,
$ \rh_{\pm} \cdot \rh_{\mp} = 0 $.
Therefore, we conclude that, in the $ \uh $--$ \nh $ algebras,
the case $ V_2 $ with $ a^2 \neq 1$ is {\it equivalent} to the case
$ I_a $, or either the $ I_b $. For $ a^2 = 1 $, one gets
\beq
{q}= 2 f \ll, \nnb
\eeq
with $ \rh_{\pm} $ a null 1-form. Therefore, we conclude
that the case $ V_1 $ with $ a^2 = 1 $ is equivalent to a
Kerr-Schild-like problem.

We know that $ \{ \rh_{+}, \rh_{-} \} $ is
a combination of $ \uh $ and $ \nh $. One can check, using
the expressions of $ \uh' $, $ \nh' $ given above, that after
some algebra $ \{ \rh'_{+}, \rh'_{-} \} $ verifies the
transformation law that corresponds to each case. For $ a^2 = 1 $,
$ \alpha_0 $ is no longer fixed by the
relations~\rf{eq22}. Thus, $ \alpha_0 $
adds the {\it fourth} degree
of freedom which characterizes \ks groups.
This constitutes an important confirmation of
the coherence of the whole scheme.
\section{A procedure to find all \ks motions in flat spacetime}
\label{app-d}

Let us now begin with $ \lh $ non-geodesic
and $ \lh $ geodesic with  $ \Delta \neq 0 $. In both situations
the \ks problem reduces
to a problem of {\em restricted isometries} in flat spacetime. The point is
that one can then use a classification of all subgroups of the Poincar\'e
group to find all the \ks motions of this type.

One should focus on the restricting equation, i.e.,
$ \lie \lh = m \lh $, or equivalently $ \lie \lv = m \lv $.
Commuting each Killing vector field with a general $ \lv
$
of flat spacetime, we get explicit conditions on the functional dependence
of
its
components. Of course, one must add the conditions that $ \lh $ be
non-geodesic,
or
geodesic with $ \Delta \neq 0 $.

We have begun to study the non-geodesic case. We have, at the
time we write this, only studied the subgroups made of either boosts,
rotations
or translations, and direct combinations of them. The main result is that at
most
there exist fourth dimensional \ks motions (restricted isometries). For
instance,
the $ G_{KS,4} $  ($ G_{KS,n} $ stands for a group of \ks motions of
dimension $ n $) are represented by Ex.~\ref{ex1}; the $ G_{KS,3} $ are
formed by three independent
translations, or either by a rotation or boost with their two possible
orthogonal translations. The $ G_{KS,2} $ and $ G_{KS,1} $ are still
simpler.
There are
no more $ G_{KS} $ in this subset of all subgroups of the Poincar\'e
group (for instance, four translations lead to a covariantly constant null
vector and the same is true for three rotations or boosts).

We will not write here the functional expression of the null
vectors that satisfy each of the conditions. They can be calculated assuming
a
general null $ \lh $ and imposing the corresponding commuting restriction.
This study has to be extended to all possible subgroups of the Poincar\'e
group. In this sense holonomy theory may be helpful (see references in
Sect.~\ref{intro}).

Finally, the same calculation should be carried out for the geodesic $ \lh $
with $ \Delta \neq 0 $ (see Sect.~\ref{s-gld0}). Were the second set of
conditions of $ \Delta = 0 $ in
flat
spacetime to lead to the cylindrical and parallel cases (see I and
Sect.~\ref{sss-sld0}), its solution could
already be read from Ex.~\ref{exlm}, Eqs.~\rf{eq-exks},~\rf{eq-hfg}. Otherwise, one
only
ought to solve the new $ \lh $ in the same way as in
that example.

\end{document}